\documentclass[iop,appendixfloats]{emulateapj}

\shorttitle{LG monitoring. II}
\shortauthors{Saremi et al.}
\usepackage[usenames,dvipsnames]{color}
\usepackage{amsmath}
\usepackage{float}
\usepackage{ctable} 
\usepackage{graphicx}
\usepackage{natbib}
\begin{document}
                                
\title{The Isaac Newton Telescope monitoring survey of Local Group dwarf galaxies. II. 
The star formation history of Andromeda I derived from long period variables}

\author{Elham Saremi\altaffilmark{1},
Atefeh Javadi\altaffilmark{1}, 
Mahdieh Navabi\altaffilmark{1},
Jacco~Th.~van~Loon\altaffilmark{2},
Habib G.~Khosroshahi\altaffilmark{1,3},
Behzad Bojnordi~Arbab\altaffilmark{3,4},
Iain McDonald\altaffilmark{5}}

\altaffiltext{1}{School of Astronomy, Institute for Research in Fundamental Sciences (IPM), P.O. Box 1956836613, Tehran, Iran; saremi@ipm.ir}
\altaffiltext{2}{Lennard-Jones Laboratories, Keele University, ST5 5BG, UK}
\altaffiltext{3}{Iranian National Observatory, Institute for Research in Fundamental Sciences (IPM), Tehran, Iran}
\altaffiltext{4}{Physics Department, Sharif University of Tecnology, Tehran 1458889694, Iran}
\altaffiltext{5}{Jodrell Bank Centre for Astrophysics, Alan Turing Building, University of Manchester, M13 9PL, UK}

%
\begin{abstract}

An optical monitoring survey in the nearby dwarf galaxies was carried out with the 2.5-m Isaac Newton Telescope (INT). 55 dwarf galaxies and four isolated globular clusters in the Local Group (LG) were observed with the Wide Field Camera (WFC). The main aims of this survey are to identify the most evolved asymptotic giant branch (AGB) stars and red supergiants at the end-point of their evolution based on their pulsational instability, use their distribution over luminosity to reconstruct the star formation history (SFH), quantify the dust production and mass loss from modelling the multi-wavelength spectral energy distributions, and relate this to luminosity and radius variations. 
In this second of a series of papers, we present the methodology used to estimate SFH based on long-period variable (LPV) stars and then derive it for Andromeda\,I (And\,I) dwarf galaxy as an example of the survey. 
Using our identified 59 LPV candidates within two half-light radii of And\,I and Padova stellar evolution models, we estimated the SFH of this galaxy. A major epoch of star formation occurred in And\,I peaking around 6.6 Gyr ago, reaching $0.0035\pm0.0016$ M$_\odot$ yr$^{-1}$ and only slowly declining until 1--2 Gyr ago. The presence of some dusty LPVs in this galaxy corresponds to a slight increase in recent star formation peaking around 800 Myr ago. We evaluate a quenching time around 4 Gyr ago ($z<0.5$), which makes And\,I a late-quenching dSph.
A total stellar mass $(16\pm7)\times10^6$ M$_\odot$ is calculated within two half-light radii of And\,I for a constant metallicity $Z=0.0007$.

\end{abstract}

%
\section{Introduction}

Dwarf galaxies in the Local Group (LG) are excellent laboratories to understand the connection between stellar populations and galaxy evolution. Often they can be classified into early-type dwarf spheroidals (dSphs) that are satellites of the Andromeda (M\,31) galaxy and Milky Way (MW), and late-type dwarf irregulars (dIrrs) at farther distances from these two galaxies. Whereas dSphs are often assumed to have exclusively old populations, some possess a dominant intermediate age population due to multiple star formation episodes. Similarly, although all dIrrs are considered young, there are a few dIrrs with intermediate age and old star populations. The star formation history (SFH) is one of the most powerful tracers of the formation and evolution of these dwarf galaxies (Wyder 2001). 

Galaxy interactions and mergers play an important role in compressing gas and triggering star formation in galaxies, and thus their chemical evolution. Even though the major mergers are violent and more effective in this sense, minor mergers and interactions between satellites and host galaxy, are expected to be more common (Bournaud 2010). MW halo substructures or stellar streams are confirmations of these minor satellite encounters. Recently, Ruiz-Lara et al.\ (2020) studied the detailed SFH of the $\sim 2$ kpc around the Sun based on the Gaia survey (e.g., Gaia Collaboration 2018) and also, the orbit of Sagittarius dwarf galaxy (Sgr). They found that episodes of enhanced star formation in the MW coincided with proposed Sgr pericentre passages according to orbit simulations and Sgr stellar content. Therefore dwarf satellites can be effective in the build-up of the host galaxy stellar mass and its internal dynamics (Ruiz-Lara et al.\ 2020).

Despite recent studies of LG dwarf galaxies' SFHs (e.g., Weisz et al.\ 2014a;  2015; Skillman et al.\ 2017), many open questions remain (Saremi et al.\ 2020). Overall, extended periods of star formation over the age of the Universe are seen in these galaxies. Mateo (1998) studied a number of LG dwarf galaxies and concluded that although they have diverse star formation, trends do exist. Weisz et al.\ (2014a) estimated that dwarf galaxies with $M<10^5$ M$_\odot$ formed more than 80\% of their stellar mass $>10$ Gyr ago (prior to redshift 2), while those with $M>10^7$ M$_\odot$ had experienced only 30\% of their overall star formation by that time. Caldwell et al.\ (2017) investigated the use of globular clusters (GCs) as luminous tracers of the SFHs of LG dwarf galaxies. While the presence of GCs indicates active early star formation in dwarf galaxies, they have not been found in most of them, especially among the less massive galaxies.
SFHs of the M\,31 dSphs are more uniform than the MW dSphs (Skillman et al.\ 2017). Based on a study of the red horizontal branches (HBs) or red clumps in some dwarf galaxies, Martin et al.\ (2017) confirmed that a large fraction of M\,31 satellites have extended SFHs and thus they do not appear to have quenched quickly with early star formation episodes. They found red HBs even in the faintest M\,31 dSphs, in contrast to MW low-mass dSphs that have more prominent blue HBs than red HBs.

Star formation in the vast majority of LG satellites has all but ceased. Various reasons are suggested for quenching, among which environmental effects are the most important mechanisms (Weisz et al.\ 2014a; Fillingham et al.\ 2015). Weisz et al.\ (2014b) investigated the signatures of reionization in the SFHs of 38 LG dwarf galaxies but they found only a few quenched galaxies that had formed the bulk of their stellar mass before reionization ($>12.9$ Gyr ago). Although the reionization did not quench most dwarf galaxies, its role in the decline of star formation cannot be ignored (Skillman et al.\ 2017). Gatto et al.\ (2013) introduced ram pressure as a dominant quenching mechanism in many MW dSphs as the hot halo can efficiently strip gas from dwarfs at a distance around 90 kpc. Besides, tidal effects alongside ram pressure has been demonstrated to help quench satellites (e.g., {\L}okas et al.\ 2012; Weisz et al.\ 2014a).

Based on observational constraints, different methods have been developed to determine the SFHs of galaxies. In resolved galaxies, the distribution of different stellar populations on colour--magnitude diagrams (CMD) is very informative, however, a high quality CMD is needed. In this regard, evolved stars such as asymptotic giant branch (AGB) stars and red supergiants (RSGs), with high luminosities, are among the best accessible tracers of stellar populations.

AGB stars and RSGs are unique objects in many aspects. They help to reconstruct the SFH of a galaxy, tracing stellar populations from as recently formed as 10 Myr ago to more ancient than 10 Gyr, because they are in the final stages of their evolution and hence their luminosity is more directly related to their birth mass (Javadi et al.\ 2011b). They are also among the most important sources of dust in the interstellar medium (ISM; Boyer et al.\ 2009; Javadi et al.\ 2013), and chemically enrich the ISM with material dredged-up from their stellar cores (notably Li, C, N, F and $s$-process elements; e.g., Karakas \& Lattanzio 2014). They are possible sources of dust in the early Universe as inferred from the spectral energy distribution of high-redshift galaxies and quasars, but this depends on their evolutionary properties in the low metallicity regime: DUSTiNGS (DUST in Nearby Galaxies with Spitzer) finds that AGB stars are efficient dust producers even down to 0.6\% Z$_\odot$ (Boyer et al.\ 2015a). They are easily detectable due to long-period variability and are widely distributed (Javadi et al.\ 2011a; 2015).

We conducted an optical survey of nearby galaxies (the most complete sample so far) with the 2.5-m Isaac Newton Telescope (INT) over ten epochs. Our main objectives include: identify all long period variable (LPV) stars in the dwarf galaxies of the LG accessible in the Northern hemisphere, and then determine the SFHs from their luminosity distributions; obtain accurate time-averaged photometry for all LPVs; obtain the pulsation amplitudes of them; determine their radius variations; model their spectral energy distributions (SEDs); and study their mass loss as a function of stellar properties such as mass, luminosity, metallicity, and pulsation amplitude.

Paper\,I presented the status of the monitoring survey of LG dwarf galaxies along with the identification of LPV stars in And\,I (Saremi et al.\ 2020; hereafter Paper\,I). And\,I is a bright dSph ($M_V= -11.7\pm0.1$ mag) within the virial radius of its host galaxy (McConnachie 2012) that was initially discovered on photographic plates by van den Bergh (1972). The distance to And\,I has been determined with several methods; we calculated a distance modulus of $24.41\pm0.05$ mag based on the tip of the RGB (Paper\,I). Also, a half-light radius of $3.2\pm0.3$ arcmin was estimated in Paper\,I by fitting the radial surface brightness profile with an exponential law.

This is Paper\,II in the series, describing the SFH of And\,I dwarf galaxy to introduce our methodology of SFH reconstruction using LPV stars. Paper\,III in the series will discuss the mass-loss mechanism and dust production rate of And\,I, and other individual and populations of dwarf galaxies will be subject of subsequent papers.

This paper is organized as follows: In Sec.\ 2, we describe the catalogue of LPV candidates belonging to And\,I. The methodology used to evaluate the SFH of our sample is outlined in Sec.\ 3. The SFH of And\,I itself is presented and discussed in Sec.\ 4. In Sec.\ 5, we discuss the cumulative SFH and quenching time of And\,I and the relation between dark matter haloes and SFH. Our conclusions are presented in Sec.\ 6.

%
\section{Data}

In this section, we briefly describe pertinent features of the LPV catalogue of And\,I produced in Paper\,I and a number of catalogues from the literature that we also used.

\subsection{LPV candidates in the INT catalogue}
\begin{figure}
\includegraphics[width=1.0\columnwidth]{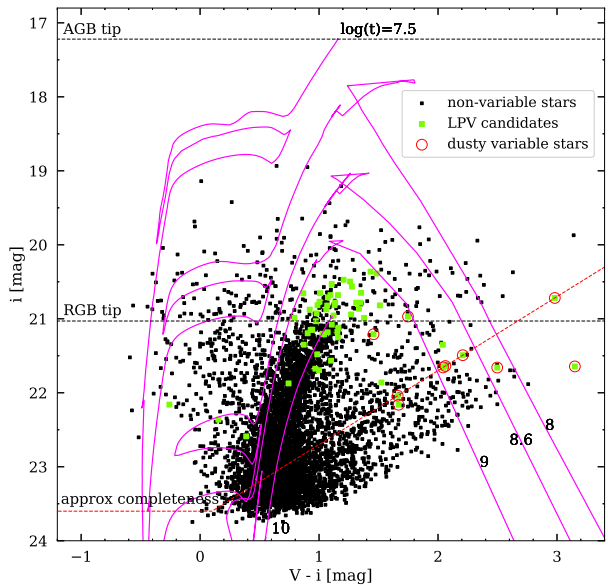}
\caption{CMD of And\,I in the $i$ band {\it vs}.\ $V-i$ colour showing our identified LPV candidates. Overplotted are isochrones from Marigo et al.\ (2017) for a distance modulus of 24.41 mag and metallicity of $Z=0.0007$. The red open circles show some dusty LPV candidates which have colours that were estimated based on the completeness limit. Red dashed line and black dashed lines represent the estimated completeness limit and the tips of the RGB and AGB, respectively (see Paper\,I).}
\end{figure}

In Paper\,I, we described the status of the monitoring survey of 55 LG dwarf galaxies with WFC at the INT in the $i$- and $V$-band filters (Saremi et al.\ 2017; Parto et al.\ 2021). Ten epochs of observation were taken over the period 2015–2018. By employing the Point Spread Function (PSF) fitting method (the {\sc daophot/allstar/allframe} software; Stetson 1987), photometry was obtained for all stars in our crowded stellar fields. 
Also, we presented a photometric catalogue of And\,I as a first result of our survey in Paper\,I. A search for large-amplitude LPV candidates was performed in the region of CCD 4 of WFC ($11.26\times22.55$ arcmin$^2$) with central coordinate $00^{\rm h}45^{\rm m}39\rlap{.}^{\rm s}9$, 
$+38^\circ02^\prime28^{\prime\prime}$ (Saremi et al.\ 2019b). We found 9\,824 stars in this region, among which 5\,581 stars were within two half-light radii of the centre of And\,I. 
The catalogue reaches a completeness limit of 100\% above the tip of the red giant branch (RGB) around $i\sim 21$ mag. Foreground stars were eliminated by applying criteria on parallax and proper motion measurements from Gaia Data Release 2 (DR2; Gaia Collaboration 2018).

By using a method similar to the {\sc newtrial} routine (Welch \& Stetson 1993; Stetson 1996), we identified 59 LPV candidates within two half-light radii of And\,I and 97 candidates across the whole of CCD 4 of WFC. The amplitudes of these candidates range from 0.2 to 3 mag in the $i$-band. Fig.\ 1 shows a CMD of And\,I in the $i$ band {\it vs}.\ $V-i$ colour showing our identified LPV candidates. Overplotted are isochrones from Marigo et al.\ (2017) for a distance modulus of 24.41 mag and metallicity of $Z=0.0007$ (cf.\ Sec.\ 3.1).
The colours of some LPV candidates (highlighted by red open circles) are estimated on the basis of the completeness limit because they are detected only once or twice in the $V$ band (or not at all). These candidates are likely to be dusty stars which were not easily seen in the $V$ band (see Paper\,I). The estimated completeness limit (cf.\ Sec.\ 3.2 of Paper\,I) is indicated with the red dashed line in Fig.\ 1; the horizontal dashed lines represent the tips of the RGB and AGB (cf.\ Sec.\ 5.1 of Paper\,I).

\begin{figure}
\includegraphics[width=1.0\columnwidth]{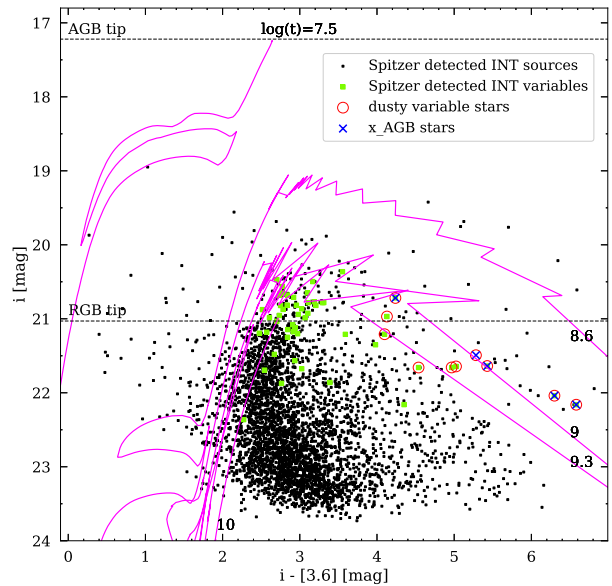}
\caption{CMD of common sources from the {\it Spitzer} catalogue and our INT survey in the $i$ band {\it vs}.\ $i-[3.6]$ colour showing our identified LPV candidates. Overplotted are isochrones from Marigo et al.\ (2017) for a distance modulus of 24.41 mag and metallicity of $Z=0.0007$. The red open circles are dusty LPV candidates. Five x-AGB stars identified in Paper\,I are highlighted with blue crosses.
}
\end{figure}

\subsection{Cross-correlation with {\it Spitzer} catalogue}

Paper\,I also describes the cross-correlation of our data with the mid-IR data from the {\it Spitzer} Space Telescope (Boyer et al.\ 2015b). Unfortunately, there is no near-IR photometry in the $J$, $H$ and $K$ bands at the necessary depth for most M\,31 dwarf galaxies. However, mid-IR magnitudes exist for $\sim 75$\% of our And\,I optical data and 98\% of LPV candidates. Fig.\ 2 shows a CMD of common sources in the $i$ band {\it vs}.\ $i-[3.6]$ colour, with INT variables highlighted in green. Similar to Fig.\ 1, the Padova isochrones are overlain in magenta and red open circles identify dusty LPV candidates. Since we do not have the precise $V-i$ colour of dusty LPV candidates, we will use their $i-[3.6]$ colour in SFH estimates.

We identified five extreme AGB (x-AGB) stars among our LPV candidates in And\,I that are brighter than $M_{3.6}$ = -8 mag with colours $[3.6]-[4.5] > 0.1$ mag (highlighted with blue crosses on Fig.\ 2; cf.\ Sec.\ 5.3 of Paper\,I). Although the x-AGB stars represent a minor population of stars (less than 6\% of the total AGB population), they make more than 75\% of the dust produced by cool evolved stars (e.g., Riebel et al.\ 2012; Boyer et al.\ 2012). Three x-AGB out of five are in common with the DUSTiNGS survey (Boyer et al.\ 2015b).

%
\section{Star Formation History}

One of the main objectives of our survey is the reconstruction of the SFH of LG dwarf galaxies using a uniform methodology in data gathering and analysis. We derive the SFH of galaxies based on the AGB (and RSG) phase of stellar evolution. When identifying LPV stars, we determine their magnitudes; theoretical stellar evolution models help us to transform LPVs' brightnesses into their birth masses. To this aim, we use the Padova model, which is one of the best available models (Marigo et al.\ 2017). This stellar model has calculated most of the evolutionary phases of AGB stars; from thermal pulsing AGBs until they enter the post-AGB phase, accounting for the effects of third dredge-up and hot-bottom burning (HBB; Iben \& Renzini 1983). The isochrones of Padova models are the most appropriate for our purpose because they predict the variability properties and the effects of circumstellar dust production on the photometry (Trabucchi et al.\ 2021; Javadi et al.\ 2011b). The LPV stars are located at the cool end of each isochrone.

%
 
\subsection{Metallicities}

Since the behaviour of the isochrones depends on metallicity, we first check the metallicities of our sample of galaxies. Fig.\ 3 shows their distribution over metallicity ([Fe/H] and $Z$ on left and right axes, respectively) {\it vs}.\ mass. The data are available in Table.\ 1 of Paper\,I. As one can see, most of the M\,31 satellites have metallicities and masses somewhat higher than MW satellites. The general relation between stellar mass and metallicity is reproduced (Kirby et al.\ 2020). The isolated dwarf galaxies naturally have higher masses and metallicities than others. The location of And\,I, highlighted with a black square on Fig.\ 3, shows this galaxy has a relatively high metallicity and mass ($> 10^6$ M$_\odot$) compared to the majority of LG satellites.

To convert the metallicities to $Z$ in this paper, we used ${\rm [M/H]}=\log(Z/X)-\log(Z/X)_\odot$, with $(Z/X)_\odot= 0.0207$ and $Y = 0.2485+1.78 Z$ obtained from the solar calibration for hydrogen, helium and metal mass fractions (Bressan et al.\ 2012). 
Salaris \& Cassisi (2005) calculated a relation ${\rm [M/H]} \sim {\rm [Fe/H]} + \log (0.694f_\alpha + 0.306)$, where $f_\alpha=10^{[\alpha/{\rm Fe}]}$. 
The ratio of $\alpha$ elements (Mg, Ca, Si, Ti) to Fe can be a good tracer for the SFH of a galaxy because they are produced on different timescales (Wojno et al.\ 2020).
Vargas et al.\ (2014) measured [Fe/H] and [$\alpha$/Fe] of stars in some of M\,31 satellites including And\,I ([Fe/H] $= - 1.109\pm0.117$ dex and [$\alpha$/Fe] $= 0.278\pm0.164$ dex). Kirby et al.\ (2020), with a higher signal to noise in their individual spectra, found lower ratios, [Fe/H] $= - 1.51\pm0.02$ dex and [$\alpha$/Fe] $= 0.14\pm0.04$ dex for this galaxy which corresponds to $Z \sim 0.0006$.
Wojno et al.\ (2020) estimated [Fe/H] $= - 1.36\pm0.004$ dex that, with [$\alpha$/Fe] $= 0.14\pm0.04$ dex, yields $Z \sim 0.0008$. In this paper, we selected $Z=0.0007$ for And\,I, however, we have also examined the SFH under the assumption of $Z=0.0006$ and $Z=0.0008$.

\begin{figure}
\includegraphics[width=1.0\columnwidth]{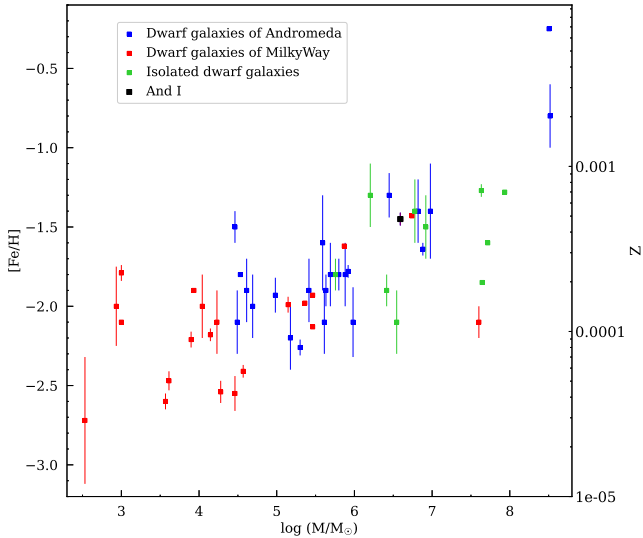}
\caption{Distribution of the survey sample of galaxies over metallicity ([Fe/H]) {\it vs}.\ mass. And\,I is highlighted with a black square.}
\end{figure}
%
\subsection{Methodology}

To reconstruct the SFH, we adapt the method previously applied to other LG dwarf galaxies (Javadi et al.\ 2011b, 2017; Rezaeikh et al.\ 2014; Hamedani Golshan et al.\ 2017; Hashemi et al.\ 2019; Saremi et al.\ 2019a); here we only provide a summary. The SFH is described by the star formation rate (SFR), $\xi$, in M$_\odot$ yr$^{-1}$ -- the rate at which gas mass was converted into stars per year. The number of stars formed can be related to the total mass of the stars created, with the help of an initial mass function (IMF). We use the IMF defined by Kroupa (2001) and the minimum and maximum of the stellar mass range are adopted to be 0.02 and 200 M$_\odot$, respectively.

With supposing the stars with mass between $m(t)$ and $m(t+dt)$ as LPV stars at present, and $\delta t$ as the ``pulsation duration'' (duration of the evolutionary phase of large-amplitude, long-period variability), finally, we will have a relation for the SFR:

\begin{equation}
\xi(t)=\frac{{\rm d}n^\prime(t)}{\delta t}\
\frac{\int_{\rm min}^{\rm max}f_{\rm IMF}(m)m\,{\rm d}m}
{\int_{m(t)}^{m(t+{\rm d}t)}f_{\rm IMF}(m)\,{\rm d}m},
\end{equation}
where ${\rm d}n^\prime$ is the number of LPV stars that we can observe in an age bin.
Thus, to derive the SFH of a galaxy, we require relations between mass and luminosity, and between age and pulsation duration for individual LPV stars.

%
\begin{figure*}
\centering
\includegraphics[width=2.1\columnwidth]{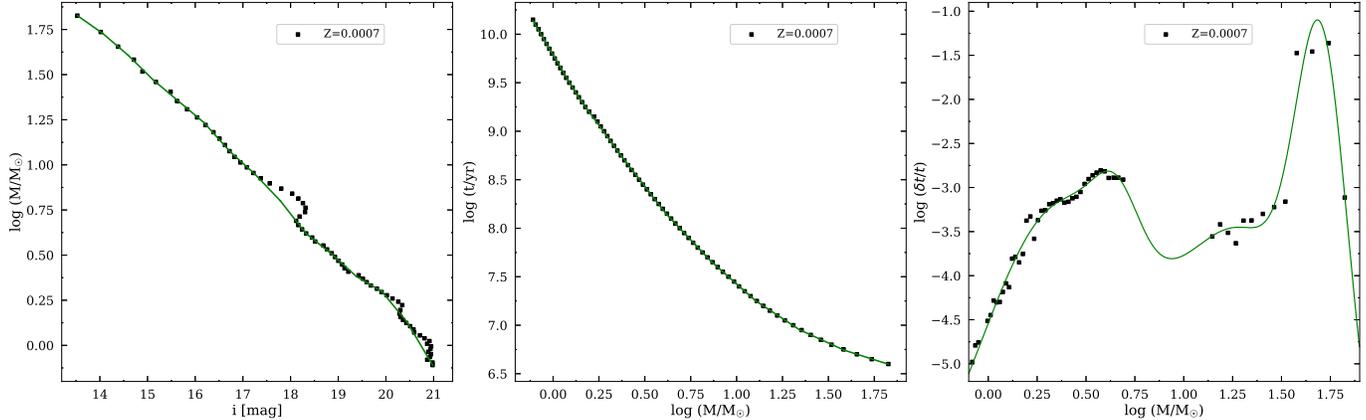}
\caption{Left panel: The mass--luminosity relation in the $i$ band for Galactic extinction $A_i=0.105$ mag, metallicity $Z=0.0007$ and distance modulus $\mu=24.41$ mag of And\,I. The linear spline fits, drawn as green solid lines are obtained by minimisation of $\chi^2$. Middle panel: The mass--age relation for a metallicity $Z=0.0007$ along with linear spline fits. Right panel: The mass--pulsation duration relation for the same metallicity, where the points show the ratio of pulsation duration to age {\it vs}.\ mass; the solid lines are multiple-Gaussian fits. All of the relations are obtained from Padova isochrones (Marigo et al.\ 2017).}
\end{figure*}

\subsubsection{Mass--Luminosity relation}

Identified LPV stars in our monitoring survey should have reached the very final stages of their evolution and thus their brightness can be translated into their birth mass. Thus, using isochrones, we can construct a mass--$i$-magnitude relation for different metallicities.
In the left panel of Fig.\ 4 is shown the mass--luminosity relation in the $i$ band for the Galactic extinction, metallicity and distance modulus of And\,I ($A_i=0.105$ mag, $Z=0.0007$ and $\mu=24.41$ mag, respectively). The coefficients of the best fitting function are listed in Table 1. The linear spline fits, drawn as green solid lines are obtained by minimisation of $\chi^2$ with the Iraf task {\sc gfit1d}. There is a little excursion towards fainter $i$-band magnitudes on the plot (around 17--18 mag), maybe because of the difficulty of modelling high-mass AGB stars and super-AGB stars ($\sim5$--10 M$_\odot$) and/or due to the change in the atmospheric composition of these stars resulting from HBB; anyway, we interpolated over that range in mass (see Javadi et al.\ 2011b). However, most of our sample of dwarf galaxies possess no stellar populations with such high masses, so this part of the plot is often irrelevant for our analysis.

All figures and tables for the range of metallicities of relevance to our sample of galaxies are presented in the Appendix; they were calculated for absolute magnitudes in the $i$ band and thus the distance modulus and the Galactic extinction of each galaxy must be applied to them.


Before using the mass--luminosity relation to convert the $i$-band magnitudes of the LPV stars to their masses, it is necessary to apply extinction corrections for stars that are dimmed and reddened by circumstellar dust. Interstellar dust -- that is different in optical properties -- is applied as Galactic extinction in the relations. Although the isochrones of Marigo et al.\ (2017) have considered the photometric effects of circumstellar dust to some extent, also, we need to apply a de-reddening correction for dusty LPV stars. 
To do this, we traced the reddened stars on the isochrones with a range of ages in Fig.\ 1. The LPV candidates that are redder than the peak of each isochrone must be returned to the peak. Often reddening for carbon-rich stars (C stars) occurs faster than for oxygen-rich stars (M-type stars); thus we calculated two average slopes of extinction, $a=1.37$ mag mag$^{-1}$ for C stars (e.g., at $t=1$ Gyr, $\log t=9$) and $a=2.04$ mag mag$^{-1}$ for M-type stars (e.g., at $t=100$ Myr, $\log t=8$). The reddening correction will be applied to stars with $(V-i)>1.4$ mag as:
\begin{equation}
i _{\rm cor} = i+a[(1.16)-(V-i)],
\end{equation}
where the peak of isochrones is estimated in colour of $(V-i)\sim1.16$ mag. 

As mentioned in Sec.\ 2.1, some dusty LPV candidates (highlighted by red open circles in Figs.\ 1 and 2) do not have a reliable measurement of $V-i$. Hence, we used the $i-[3.6]$ colour to determine their intrinsic $i$-band magnitude. Based on Fig.\ 2, we obtained the correction equation for these stars as:
\begin{equation}
i _{\rm cor} = i+a[(2.5)-(i-[3.6])].
\end{equation}
with average slope, $a=0.32$ mag mag$^{-1}$ and the peak of isochrones in colour of $(i-[3.6])\sim2.5$ mag. All our dusty LPV candidates are C-star candidates.

Since the majority of our LPV candidates do not have observational information to distinguish C stars from M-type stars, we have to select a mass range for C stars based on theory and observations in other galaxies, notably the Magellanic Clouds (MCs). Renzini \& Voli (1981) estimated for a metallicity of $Z=0.004$ that AGB stars with a minimum birth mass of 1.1 M$_\odot$ can become a C star. Groenewegen \& de Jong (1993) conducted a population synthesis study of AGB stars in the Large Magellanic Cloud (LMC) and found that stars with $M < 1.2$ M$_\odot$ do not experience third dredge-up and thus remain of M type. Stars with birth masses between 1.3 and 4 M$_\odot$ were identified to be C stars by van Loon, Marshall \& Zijlstra (2005); they obtained this limit based on clusters in the MCs. With the same method, Girardi \& Marigo (2007) estimated a range of $1.5 < M < 2.8$ M$_\odot$ for C stars in some MC clusters. 
But these estimates are valid for solar and slightly sub-solar metallicity while most of our LPV candidates will have lower metallicities (see Fig.\ 3). The galaxies with lower metallicities initially contain less oxygen in their atmospheres, therefore, AGB stars in these galaxies will more easily become C stars (Leisenring et al.\ 2008). Thus, a range of $1.1 < M < 4$ M$_\odot$ is adopted for the birth mass of C stars.

We applied the carbon-dust correction to all reddened stars and estimated the mass of each LPV candidate; if it does not have the selected mass for a C star, we re-applied the oxygenous dust correction. In Fig.\ 5, we present the distribution over brightness and birth mass of the variable stars within two half-light radii of And\,I. In this figure, the dashed red lines are when the reddening correction has not yet been applied and green lines are after having applied it; it is clear that correcting for the effect of circumstellar dust transfers stars to brighter magnitudes and thus higher masses. There is a peak at $M \sim 1.1$ M$_\odot$ and the highest mass obtained for LPV candidates in And\,I is 2.23 M$_\odot$. Hence, as expected from dSph galaxies, we see old but also intermediate-age populations in this galaxy.

It must be stressed that for the lowest-mass populations (oldest ages), the SFR is likely to be incomplete because these stars may not exhibit large-amplitude variability. 
Based on our method, we can only estimate the correct mass for stars that have reached the final phase of evolution. The LPV candidates below the tip of the RGB, which are probably short-period variables, have not yet reached this stage. Although we estimated the mass of these LPV candidates based on the slope of the last fitted line in mass–luminosity relation (left panel of Fig.\ 4) and presented them in Fig.\ 5, their mass is too small and this makes their age older than the age of the Universe.  It should be noted that these stars have no effect on our calculations generally because we estimate the SFH of a galaxy from the beginning of the Universe and thus they are automatically excluded from the conclusion.

In Paper\,I, we estimated contamination for LPV candidates of And\,I around 28\% that decreases to 13\% for younger LPV candidates ($i<21$ mag). We suggested that most of this contamination comes from the Galactic foreground and often the contamination from M\,31 is negligible.

\begin{figure}
\includegraphics[width=1.05\columnwidth]{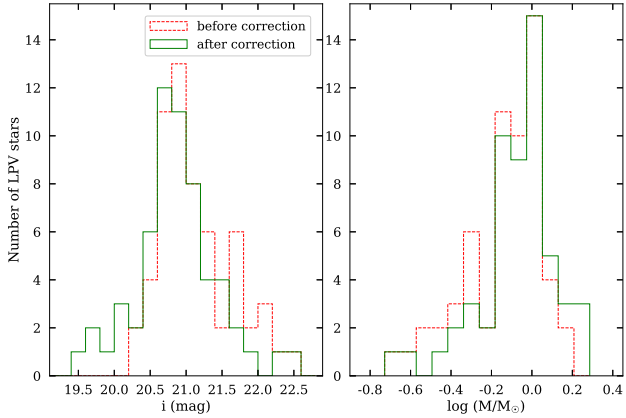}
\caption{The distribution of LPV candidates within two half-light radii of And\,I as a function of $i$-band brightness (left panel), and their birth mass function for a metallicity $Z=0.0007$ (right panel) are presented as green solid lines. Red dashed lines show the distributions before reddening correction.}
\end{figure}
\begin{table}
\caption[]{Relation between birth mass and $i$-band magnitude, $\log{M\,{\rm [M}_\odot{\rm ]}}=ai+b$, for a distance modulus of $\mu=24.41$ mag and Galactic extinction $A_i=0.105$ mag.}
\begin{tabularx}{\linewidth}{ccc}

\hline \hline \\
\hspace{0.05 in}$a$              & \hspace{0.1 in}$b$             & \hspace{0.1 in}validity range\vspace{0.05 in}\\
\hline\\
\multicolumn{3}{c}{$Z=0.0007$}\vspace{0.05 in} \\
\hline\\
\hspace{0.07 in}$ -0.186\pm0.063 $ & \hspace{0.12 in} $ 4.341\pm0.908 $ & \hspace{0.12 in} $i\leq 14.060 $ \vspace{0.04 in}\\
\hspace{0.07 in}$ -0.227\pm0.064 $ & \hspace{0.12 in} $ 4.917\pm0.953 $ & \hspace{0.12 in} $ 14.060 <i\leq 14.592 $ \vspace{0.04 in}\\
\hspace{0.07 in}$ -0.260\pm0.062 $ & \hspace{0.12 in} $ 5.402\pm0.958 $ & \hspace{0.12 in} $ 14.592 <i\leq 15.125 $ \vspace{0.04 in}\\
\hspace{0.07 in}$ -0.221\pm0.055 $ & \hspace{0.12 in} $ 4.813\pm0.868 $ & \hspace{0.12 in} $ 15.125 <i\leq 15.657 $ \vspace{0.04 in}\\
\hspace{0.07 in}$ -0.223\pm0.047 $ & \hspace{0.12 in} $ 4.838\pm0.776 $ & \hspace{0.12 in} $ 15.657 <i\leq 16.189 $ \vspace{0.04 in}\\
\hspace{0.07 in}$ -0.296\pm0.042 $ & \hspace{0.12 in} $ 6.019\pm0.705 $ & \hspace{0.12 in} $ 16.189 <i\leq 16.721 $ \vspace{0.04 in}\\
\hspace{0.07 in}$ -0.239\pm0.042 $ & \hspace{0.12 in} $ 5.071\pm0.728 $ & \hspace{0.12 in} $ 16.721 <i\leq 17.253 $ \vspace{0.04 in}\\
\hspace{0.07 in}$ -0.277\pm0.094 $ & \hspace{0.12 in} $ 5.736\pm1.703 $ & \hspace{0.12 in} $ 17.253 <i\leq 17.786 $ \vspace{0.04 in}\\
\hspace{0.07 in}$ -0.338\pm0.091 $ & \hspace{0.12 in} $ 6.816\pm1.679 $ & \hspace{0.12 in} $ 17.786 <i\leq 18.318 $ \vspace{0.04 in}\\
\hspace{0.07 in}$ -0.210\pm0.037 $ & \hspace{0.12 in} $ 4.473\pm0.703 $ & \hspace{0.12 in} $ 18.318 <i\leq 18.850 $ \vspace{0.04 in}\\
\hspace{0.07 in}$ -0.240\pm0.037 $ & \hspace{0.12 in} $ 5.027\pm0.719 $ & \hspace{0.12 in} $ 18.850 <i\leq 19.382 $ \vspace{0.04 in}\\
\hspace{0.07 in}$ -0.155\pm0.039 $ & \hspace{0.12 in} $ 3.383\pm0.795 $ & \hspace{0.12 in} $ 19.382 <i\leq 19.915 $ \vspace{0.04 in}\\
\hspace{0.07 in}$ -0.322\pm0.034 $ & \hspace{0.12 in} $ 6.704\pm0.707 $ & \hspace{0.12 in} $ 19.915 <i\leq 20.447 $ \vspace{0.04 in}\\
\hspace{0.07 in}$ -0.382\pm0.027 $ & \hspace{0.12 in} $ 7.928\pm0.580 $ & \hspace{0.12 in} $i> 20.447 $\vspace{0.05 in}\\
\hline\\
\end{tabularx}
\end{table}

%
\subsubsection{Age--Mass relation}

The middle panel of Fig.\ 4 shows the age--mass relation for the metallicity adopted for And\,I, based on Padova models. There is a piece-wise linear relation between the birth mass of LPV stars and their age at present, $\log t=a\log{M\,{\rm [M}_\odot{\rm ]}}+b$. The solid green lines are linear spline fits with coefficients listed in Table 2. Other figures and tables of age--mass relations for the range of metallicities relevant to our sample of galaxies are presented in the Appendix.
\begin{table}
\caption[]{Relation between age and birth mass, $\log t=a\log{M\,{\rm [M}_\odot{\rm ]}}+b$.}
\begin{tabularx}{\linewidth}{ccc}
\hline \hline\\
\hspace{0.05 in}$a$              & \hspace{0.09 in}$b$             & \hspace{0.1 in}validity range\vspace{0.05 in}\\
\hline\\
\multicolumn{3}{c}{$Z=0.0007$} \vspace{0.05 in}\\
\hline\\
\hspace{0.07 in}$ -3.189\pm0.024 $ & \hspace{0.09 in} $ 9.788\pm0.006 $ & \hspace{0.1 in} $\log M\leq 0.133 $\vspace{0.04 in}\\
\hspace{0.07 in}$ -2.594\pm0.022 $ & \hspace{0.09 in} $ 9.709\pm0.011 $ & \hspace{0.1 in} $ 0.133 <\log M\leq 0.375 $\vspace{0.04 in}\\
\hspace{0.07 in}$ -2.441\pm0.023 $ & \hspace{0.09 in} $ 9.652\pm0.017 $ & \hspace{0.1 in} $ 0.375 <\log M\leq 0.617 $\vspace{0.04 in}\\
\hspace{0.07 in}$ -2.002\pm0.025 $ & \hspace{0.09 in} $ 9.382\pm0.025 $ & \hspace{0.1 in} $ 0.617 <\log M\leq 0.859 $\vspace{0.04 in}\\
\hspace{0.07 in}$ -1.680\pm0.028 $ & \hspace{0.09 in} $ 9.105\pm0.034 $ & \hspace{0.1 in} $ 0.859 <\log M\leq 1.101 $\vspace{0.04 in}\\
\hspace{0.07 in}$ -1.248\pm0.032 $ & \hspace{0.09 in} $ 8.629\pm0.047 $ & \hspace{0.1 in} $ 1.101 <\log M\leq 1.343 $\vspace{0.04 in}\\
\hspace{0.07 in}$ -0.867\pm0.037 $ & \hspace{0.09 in} $ 8.118\pm0.064 $ & \hspace{0.1 in} $ 1.343 <\log M\leq 1.585 $\vspace{0.04 in}\\
\hspace{0.07 in}$ -0.601\pm0.045 $ & \hspace{0.09 in} $ 7.696\pm0.088 $ & \hspace{0.1 in} $\log M> 1.585 $\vspace{0.05 in}\\
\hline\\
\end{tabularx}
\end{table}
%
\subsubsection{Pulsation-duration--mass relation}

Since LPV stars do not spend the same lifetime in the pulsation phase (massive stars spend less time in the LPV phase than low-mass stars do), the calculation of pulsation duration is required. The relative pulsation duration (${\delta t}/t$) is derived for a range in mass based on Padova models for a metallicity of $Z = 0.0007$ (right panel of Fig.\ 4). Although we used the most recent Padova models (Marigo et al.\ 2017), since the variability of massive stars ($\log M/{\rm M}_\odot > 0.8$) was not considered we calculated the pulsation duration of only these stars using the previous release of the models (Marigo et al.\ 2008). However, as mentioned before, we rarely have such high-mass LPV candidates in our sample of galaxies. Multiple Gaussian functions were fit to the pulsation-duration--mass relation. Their coefficients are listed in Table 3. Figures and tables for other metallicities can be found in the Appendix.

\begin{table}
\caption[]{Relation between relative pulsation duration and birth mass, $\log(\delta t/t)= D + \Sigma_{i=1}^4a _i\exp\left[-(\log M [{\rm M}_\odot]-b_i)^2/c_i^2\right]$.}
\begin{tabularx}{\linewidth}{ccccc}
\hline\hline\\
\hspace{0.15 in} D &  \hspace{0.25 in} $i$ & \hspace{0.25 in}$a$   & \hspace{0.25 in}$b$  & \hspace{0.25 in}$c$  \vspace{0.05 in}  \\
\hline\\
    \multicolumn{5}{c}{$Z=0.0007$} \vspace{0.05 in}  \\
  \hline\\
\hspace{0.15 in} $-7.165$ &\hspace{0.25 in}1     &   \hspace{0.25 in}$3.392$  & \hspace{0.25 in}$1.330$  & \hspace{0.25 in}$0.497$\vspace{0.04 in}\\
                        &\hspace{0.25 in}2     &   \hspace{0.25 in}$4.076$  & \hspace{0.25 in}$1.710$  & \hspace{0.25 in}$0.185$\vspace{0.04 in}\\
                        &\hspace{0.25 in}3     &   \hspace{0.25 in}$0.707$  & \hspace{0.25 in}$0.657$  & \hspace{0.25 in}$0.163$\vspace{0.04 in}\\
                        &\hspace{0.25 in}4     &   \hspace{0.25 in}$3.903$  & \hspace{0.25 in}$0.368$  & \hspace{0.25 in}$0.583$\vspace{0.05 in}\\
\hline\\    
\end{tabularx}
\end{table}      
%
\subsection{Deriving the SFH}

To calculate the SFH, after correction for reddening (if necessary), we convert the $i$-band magnitude of each LPV candidate to its birth mass, and then we obtain its age based on the birth mass (using Tables 1 and 2). Based on Table 3, we determine the pulsation duration and then, by inverting it, we assign a weight to each star. Then, we solve equation 1 to derive the SFR in each age bin. To have uniform uncertainties in the SFR values, we first arrange the stars by mass and then we select the age bin size such that they have equal numbers of stars. The number of old, low-mass LPV candidates is much larger than the number of young, massive LPV candidates. A statistical error is calculated from the number of stars in each age bin, $N$, assuming Poisson statistics, as follows:
\begin{equation}
\sigma_{\xi(t)}\ =\ \frac{\sqrt{N}}{N}\ \xi(t).
\end{equation}
Finally, we apply the IMF correction based on the minimum and maximum mass of each bin.

%
\section{The SFH of And\,I}

Using the identified LPV candidates within two half-light radii of And\,I and applying the described method, we obtained the SFR as a function of look-back time (age) in this galaxy. Fig.\ 6 shows the SFH for a constant metallicity of $Z=0.0007$ (in black), over a time interval from 600 Myr to 13.8 Gyr ago. The spread in age within each bin and the statistical errors are shown in the horizontal and vertical ``error bars'', respectively. 
A major epoch of star formation occurred in And\,I, peaking around 6.6 Gyr ago ($\log t=9.82$) at $0.0035\pm0.0016$ M$_\odot$ yr$^{-1}$, which slowly declined until 1--2 Gyr ago. Of the total stellar mass within two half-light radii of And\,I, $(16\pm7)\times10^6$ M$_\odot$ (see Table 4), 60\% formed between 11.6 and 5.5 Gyr ago.

The recent increase in SFR, peaking around 790 Myr ago (considering the errors, 1.1 Gyr to 630 Myr ago) is very interesting. As mentioned before, there are some dusty LPV candidates in this galaxy, five of which are x-AGB stars. All of these dusty LPV candidates (highlighted in red open circles in Figs.\ 1 and 2) should have masses around 1.1 to 2.5 M$_\odot$ (ages about 4.8 Gyr to 630 Myr) and thus they are C stars (cf.\ Sec.\ 3.2.1). Our x-AGB stars (highlighted in blue crosses in Fig.\ 2) range in ages from 660 Myr to 1.4 Gyr. In fact, these stars are the tracers of the last epoch of star formation. The presence of these dusty LPV stars and especially x-AGB stars in And\,I leads us to assume that maybe there have also been such dusty stars in And\,I in previous epochs, which may have led to star formation by returning mass to the galaxy. However, more research is needed to corroborate the idea that this galaxy has been rejuvenated and to what extent stellar death has been able to replenish the interstellar medium (ISM) with metals and dust. In either case, recent star formation and young stars have been reported in relatively low-mass quenched galaxies (e.g., de Lorenzo-C\'aceres et al.\ 2020).

\begin{figure}
\includegraphics[width=1.0\columnwidth]{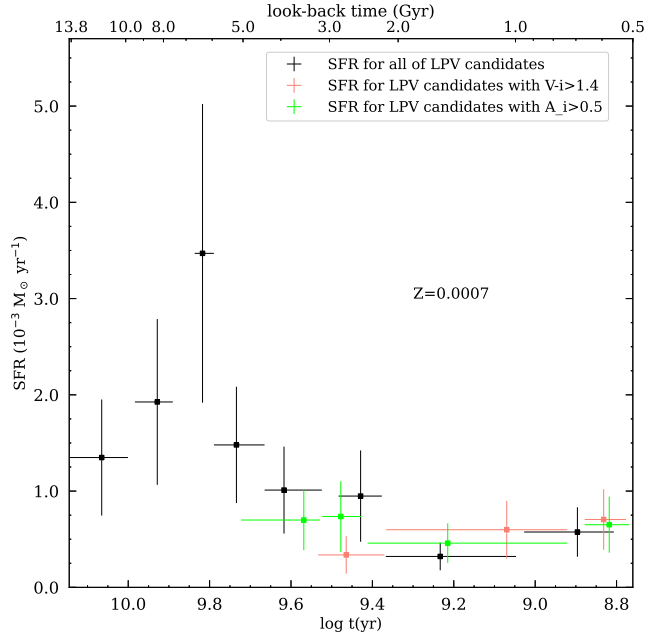}
\caption{SFH of And\,I for constant metallicity of $Z=0.0007$. The horizontal ``error bars'' show spread in age within each bin and the statistical errors are presented by vertical ``error bars''. SFH derived for only LPV candidates with $(V-i)>1.4$ mag (in salmon) and amplitude of $A_i>0.5$ mag (in light green) are shown for comparison with the original result (in black).}
\end{figure}

Da Costa et al.\ (1996) studied the SFH of And\,I using the {\it Hubble} Space Telescope/Wide Field Planetary Camera 2 (HST/WFPC2) and found evidence for an extended period of star formation, like in many of the Galactic dSphs. They estimated that the bulk of the stellar population in And\,I is $\sim 9.5\pm2.5$ Gyr old. In analysing the CMD, they confirmed the presence of red HB, blue HB and RR\,Lyr{\ae} stars in this galaxy, representing an older population than the bulk of the And\,I population; thus they estimated a comparable age to the GCs of the inner Galactic halo for And\,I. In confirmation of this, Grebel et al.\ (2000) discovered a faint GC $\sim207$ pc from the centre of And\,I, despite GCs being rare among LG galaxies with low stellar masses (Caldwell et al.\ 2017).

Martin et al.\ (2017) also showed the existence of red and blue HB stars in a large fraction of M\,31 dwarf galaxies, which indicates extended SFHs in them because of the age-sensitivity of HB stars. While blue HB stars represent old populations, red HBs indicate the presence of younger stars. Thus they reject early star formation episodes that rapidly quenched these galaxies. 
Skillman et al.\ (2017) studied the HST images of six M\,31 dSphs, including And\,I, to determine whether the SFHs of the M\,31 and MW dSph satellites are similar. They confirmed the presence of the blue and red HB populations in And\,I. Also, they observed a significant blue plume in the CMD below the HB and above the oldest main-sequence turn-off (MSTO) that resembled a ``blue straggler'' population. One scenario envisages these stars originating from close collisional or primordial binaries (e.g., Monelli et al.\ 2012) and so the presence of ``blue straggler'' stars might indicate the possibility of rejuvenation in a galaxy's evolution.
Helium-burning stars are also seen in CMDs from Skillman et al.\ (2017). These stars, with ages between $\sim2$--4 Gyr, usually are accompanied by a population of bright main-sequence stars, however, they did not see these stars in their CMDs.

\begin{figure}
\includegraphics[width=1.0\columnwidth]{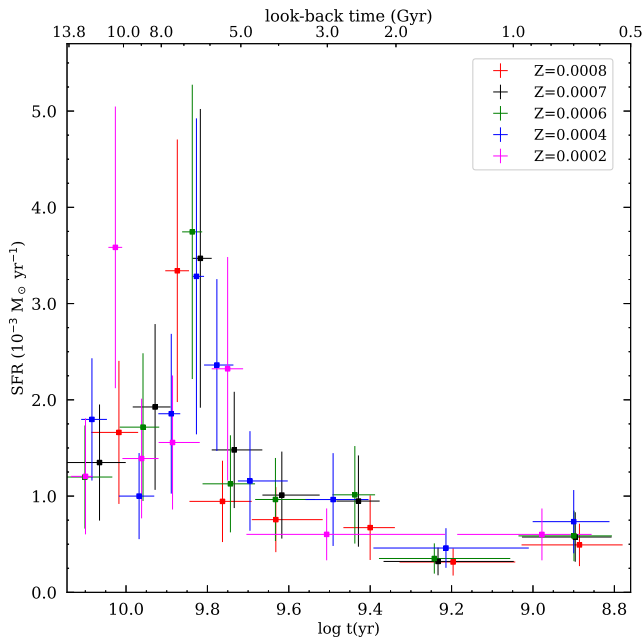}
\caption{SFHs of And\,I for five choices of metallicities: $Z=0.0008$ (red), $Z=0.0007$ (black), $Z=0.0006$ (green), $Z=0.0004$ (blue) and $Z=0.0002$ (magenta).}
\end{figure}

We estimated the total stellar mass formed (and currently) inside the half-light radius of And\,I to be $(5.9\pm2.7)\times10^6$ M$_\odot$ (see Table 4), which is in good agreement with the literature results. Woo et al.\ (2008) estimated $M_\star = 7.1\times10^6$ M$_\odot$ based on the histogram of relative SFR as a function of age created by Mateo (1998). McConnachie (2012) calculated a total stellar mass of $3.9\times10^6$ M$_\odot$ for this galaxy. 
A total stellar mass of $(7.2\pm2.5)\times10^6$ M$_\odot$ was obtained by Kirby et al.\ (2020) by using the luminosity of And\,I (Tollerud et al.\ 2012) and a stellar mass to light ratio of 1.6 M$_\odot$/L$_\odot$ (Woo et al.\ 2008). McConnachie et al.\ (2018) assumed a stellar mass to light ratio of 1.2 M$_\odot$/L$_\odot$ (McGaugh \& Schombert 2014) and calculated $M_\star =5.7\times10^6$ M$_\odot$ for And\,I. By integrating the SFH of And\,I, Weisz et al.\ (2014a) found $M_\star = (2.8^{+2.4}_{-2.8})\times10^6$ M$_\odot$ in an area $0.24\times(\pi r_{\rm h}^2)$. In the same area for comparision, we calculated a stellar mass of $(2.5\pm1.7)\times10^6$ M$_\odot$, which is very close to their estimate.

In Fig.\ 6, the contribution of the reddened stars with $(V-i)>1.4$ mag to the SFH is shown in salmon. These reddened stars represent 24\% of our LPV candidates, and all needed a correction to their $i$-band magnitudes. In this case, the first epoch of star formation occurred 2.9 Gyr ago ($\log t=9.46$) and it lasted until 680 Myr ago. Also, the SFH for LPV candidates with relatively large amplitude in the $i$ band, $A_i>0.5$ mag is considered in Fig.\ 6 (in light green). The SFH of these stars that comprise 37\% of all LPV candidates is restricted to ages less than 5.2 Gyr.
LPV stars with low masses are often expected to yield low amplitudes of variability, which is why we do not see stars with large amplitude at older epochs. Although large-amplitude variability has been seen in GCs ($>11$ Gyr old) these pulsations occur only for a brief period of time (a fraction of a Myr; Lebzelter et al.\ 2005; Lebzelter \& Wood 2005). High-mass LPV stars, too, have a small amplitude of pulsation due to their high intrinsic luminosity, but we do not see high-mass stars in And\,I as no measurable star formation has occurred in the last 600 Myr.

We compared the SFHs of And\,I for five choices of metallicities: $Z=0.0008$ (red), $Z=0.0007$ (black), $Z=0.0006$ (green), $Z=0.0004$ (blue) and $Z=0.0002$ (magenta) in Fig.\ 7; the related figures and tables for these metallicities are in the Appendix. The intermediate-age populations or the young stars might be somewhat higher in metals compared to the oldest LPV candidates because they formed from (more) chemically-enriched material. Fig.\ 7 shows the peak of star formation shifts towards older ages (10.6 Gyr ago) at low metallicity $Z=0.0002$; thus we can adopt a higher metallicity at recent times and a lower metallicity at more ancient times to estimate the location of the peak. 
Overall, the SFHs are not qualitatively different for different uniform metallicities in And\,I. The total stellar mass within two half-light radii and within one half-light radius of And\,I for different metallicities can be found in Table 4. The total stellar mass does not change much with metallicity in the range of $Z=0.0002$ to $Z=0.0008$.

\begin{figure}
\includegraphics[width=1.0\columnwidth]{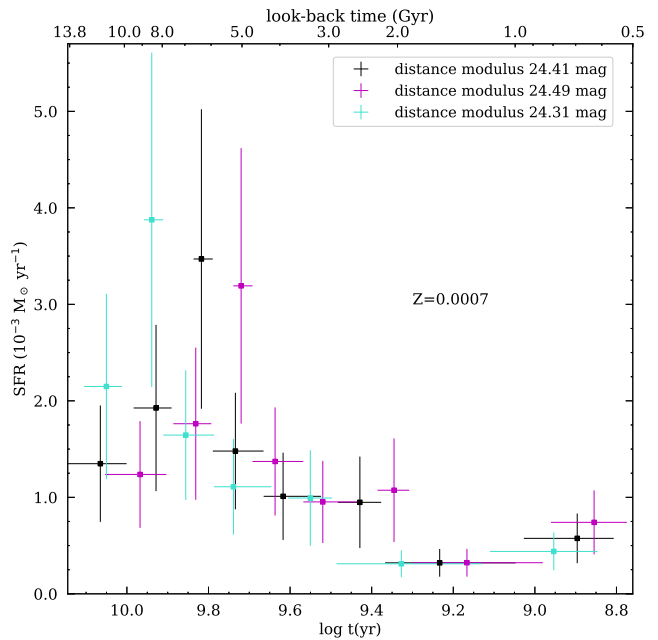}
\caption{SFHs of And\,I for constant metallicity of $Z=0.0007$ and three choices of distance modulus: $\mu=24.41$ (black; Paper\,I), $\mu=24.49$ (purple; Mart\'{\i}nez-V\'azquez et al.\ 2017) and $\mu=24.31$ mag (cyan; Conn et al.\ 2012).}
\end{figure}

In Paper\,I, we determined a distance modulus of $24.41\pm0.05$ mag to And\,I, based on the tip of the RGB. However, different distances have been obtained by other authors, for example, $\mu=24.31\pm0.07$ mag (Conn et al.\ 2012) and $\mu=24.49\pm0.12$ mag (Mart\'{\i}nez-V\'azquez et al.\ 2017). Although changing the distance modulus does not affect the age--mass and pulsation-duration--mass relations, it will change the magnitude of LPV stars and hence their mass and the SFH of the galaxy. Fig.\ 8 presents the SFH of And\,I for constant metallicity of $Z=0.0007$ for different distance moduli of $\mu=24.31$ (cyan), $\mu=24.41$ (black) and $\mu=24.49$ mag (purple) for comparison. As one can see, they have similar behaviour and there is only a shift towards recent times with increasing distance modulus.

To obtain the SFH, we used the identified LPV candidates within two half-light radii of And\,I, based on a half-light radius of $3.2\pm0.3$ arcmin that we estimated in Paper\,I. Martin et al.\ (2016) determined a half-light radius of $3.9\pm0.1$ arcmin for this galaxy after correcting for deprojection. Here, we tested the SFH for two choices of the half-light radius, 3.2 and 3.9 arcmin in black and orange, respectively (Fig.\ 9). The number of LPV candidates changes from 59 to 65 stars in the wider aperture. The SFR peak is $\sim0.004\pm0.0015$ M$_\odot$ yr$^{-1}$, slightly higher than our initial estimate, and the total mass increases to $(22\pm9)\times10^6$ M$_\odot$. Also, with this new selection for the half-light radius, the epochs of star formation would shift slightly towards more recent epochs.
\begin{figure}
\includegraphics[width=1.0\columnwidth]{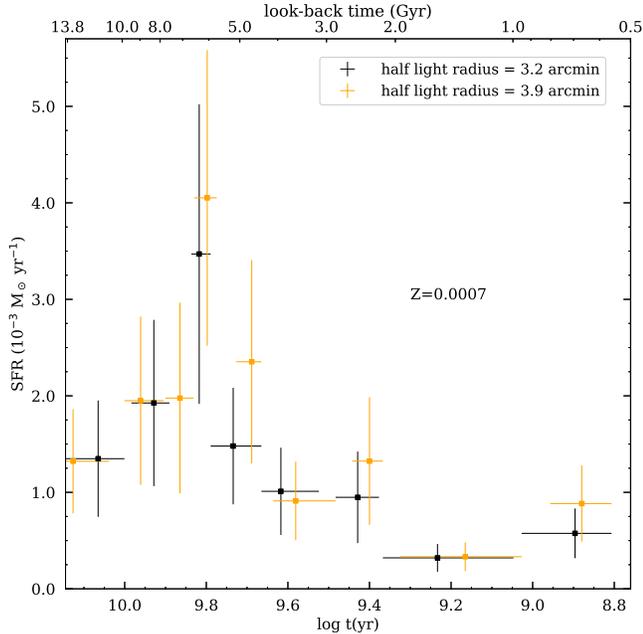}
\caption{SFHs of And\,I for constant metallicity of $Z=0.0007$ and two choices of half light radius: $r_{\rm h} = 3.2$ (black; Paper\,I) and $r_{\rm h} = 3.9$ arcmin (orange; Martin et al.\ 2016).}
\end{figure}
\begin{table}
\caption[]{Total stellar mass for different metallicities and regions.}
\begin{tabularx}{\linewidth}{ccc}
\hline \hline\\
\hspace{0.15 in}$Z$   &\hspace{0.25 in} $M_{{\rm tot},2r_{\rm h}}$ ($10^6$ M$_\odot$) & \hspace{0.25 in}$M_{{\rm tot},r_{\rm h}}$ ($10^6$ M$_\odot$) \vspace{0.05 in}\\
\hline\\
\hspace{0.15 in}0.0008& \hspace{0.3 in}$ 12\pm5  $  & \hspace{0.25 in}$ 6.5\pm3.0  $ \vspace{0.04 in}\\
\hspace{0.15 in}0.0007& \hspace{0.3 in}$ 16\pm7  $  & \hspace{0.25 in}$ 5.9\pm2.7  $ \vspace{0.04 in}\\
\hspace{0.15 in}0.0006& \hspace{0.3 in}$ 17\pm7  $  & \hspace{0.25 in}$ 6.0\pm2.8  $ \vspace{0.04 in}\\
\hspace{0.15 in}0.0004& \hspace{0.3 in}$ 14\pm6  $  & \hspace{0.25 in}$ 7.9\pm3.4  $ \vspace{0.04 in}\\
\hspace{0.15 in}0.0002& \hspace{0.3 in}$ 15\pm7  $  & \hspace{0.25 in}$ 8.5\pm4.0  $ \vspace{0.04 in}\\
\hline\\
\end{tabularx}
\end{table}

Radial population gradients have been found in some early-type dwarfs (e.g., Mateo 1998). Da Costa et al.\ (1996) suggested more centrally concentrated star formation after the initial epoch in And\,I due to its HB morphology. To distinguish whether the SFH of And\,I has varied with radius, we divided our sample into three annuli with equal numbers of LPV candidates and then we calculated the SFR density for each interval. Fig.\ 10 shows the radial gradient of SFH of And\,I for a metallicity of $Z=0.0007$. As one can see, the SFH has not varied much across different radii, only the SFR decreases with increasing radial distance. The peak of SFR is reduced by a factor $\sim15$ between the innermost and outermost regions.
\begin{figure*}
\centering
\includegraphics[width=2.1\columnwidth]{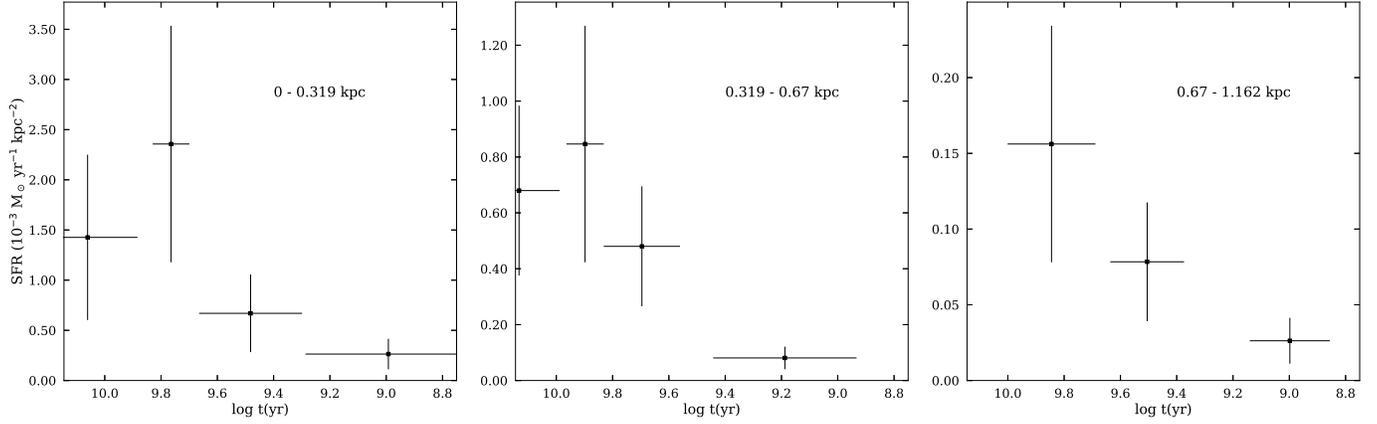}
\caption{SFR in And\,I within three regions at galactocentric radii increasing from left to right, for a metallicity of $Z=0.0007$ with equal numbers of LPV candidates in each bin.}
\end{figure*}
%
\section{Cumulative SFH and quenching time}

The cumulative SFH traces the fraction of stars formed before a given epoch; it is useful because it reduces the effects of discontinuity of time bins and small number statistics.
Fig.\ 11 shows the cumulative SFH as a function of look-back time (bottom axis) and redshift ($z$; top axis) within two half-light radii of And\,I for two metallicities $Z=0.0007$ and $0.0004$. The pink shaded area represents the error of the cumulative masses based on the statistical errors in SFRs in each bin for a metallicity of $Z=0.0007$. As expected from dSph galaxies, one can see an extended SFH with high star formation efficiency at early times. If the quenching time of a galaxy is the look-back time when 90\% of its total stellar mass is formed, then we estimate that And\,I is quenched $\sim 4$ Gyr ago ($z<0.5$).

Weisz et al.\ (2014a) used the resolved stellar populations in archival HST/WFPC2 imaging of 38 LG dwarf galaxies to derive their SFHs. Their method for estimating SFHs was the maximum likelihood CMD fitting routine, {\sc match}, first introduced by Dolphin (2002). They concluded that dSph galaxies are mainly old systems that formed most of their stars prior to $z\sim 2$ (10 Gyr ago).
They also provided the cumulative SFHs for often low-mass galaxies of the LG ($10^4 < M < 10^8$ M$_\odot$) and calculated the quenching times for them (Weisz et al.\ 2015). Using deeper CMDs, Skillman et al.\ (2017) considered the cumulative SFHs for dSphs of their sample. Table 5 and Fig.\ 12 compare our results with the results of both groups for the fraction of total stellar mass formed before the specified epoch; we reach better agreement with the Weisz et al.\ (2014a) results.

Weisz et al.\ (2015) estimated a quenching time $\log t=9.86^{+0.15}_{-0.14}$ yr ($\sim7.2^{+2.0}_{-3.0}$ Gyr ago) which is not in full accordance with our result ($\sim$ $3.7^{+0.3}_{-1.0}$ Gyr ago); although this difference is only just outside the combined error budgets. We find And\,I to be a late-quenching ($ < 5$ Gyr) dSph, contrary to the result from Skillman et al.\ (2017) and despite their observations of many young stars in their CMD (cf.\ Sec.\ 4). 
While previous determinations of the cumulative SFH for And\,I are concentrated on the limited region in the centre of galaxy, we studied two half-light radii of it (an area around 17 times larger). Furthermore, in our method, we do not need to worry about the depth of the CMD because only the completeness of AGB stars is important, which can be easily achieved (Paper\,I). In the method of counting stars on the CMD, to characterise the earliest epochs of star formation it is critical to reach the oldest MSTO. Unfortunately, HST/WFPC2 images have not reached this depth (Weisz et al.\ 2014a); though this is not an issue when studying younger populations in LG dwarf galaxies.

Ricotti \& Gnedin (2005) introduced some of the LG dSphs including And\,I as ``true fossils'' by comparing observations to simulated dwarf galaxies at $z=0$. True fossils are galaxies that formed at least 70\% of their total stellar mass before reionization and were quenched by it. It is supposed that heating of the ISM due to the high ultra-violet (UV) background in the epoch of reionization suppressed star formation in some low-mass dwarf galaxies. In Fig.\ 11, the epoch of reionization, $z \sim 6$--14 (12.9--13.5 Gyr ago; Fan et al.\ 2006) is highlighted with a purple band. We estimate that 70\% of the total stellar mass of And\,I formed in the last $\sim 6.20^{+1.57}_{-0.04}$ Gyr ago, and therefore conclude that And\,I is not a fossil galaxy. Also, Weisz et al.\ (2014b) obtained this time to be $\log t=9.88^{+0.19}_{-0.01}$ Gyr ($\sim7.59^{+0.17}_{-4.16}$ Gyr ago) for And\,I which is in accordance with our result.

\begin{figure}
\includegraphics[width=1.0\columnwidth]{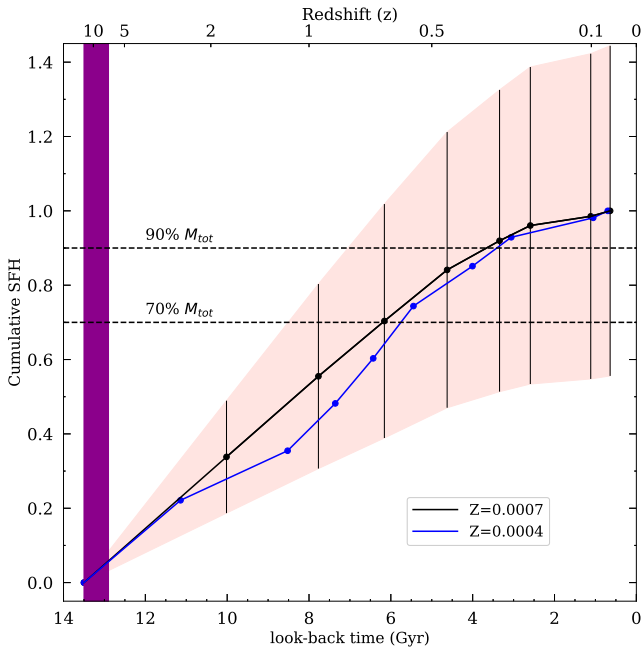}
\caption{Cumulative SFH as a function of look-back time (bottom axis) and redshift (top axis) within two half-light radii of And\,I for two metallicities: $Z=0.0007$ and $Z=0.0004$. The cumulative SFH errors for a metallicity of $Z=0.0007$ are shown with a pink shaded area based on the statistical errors in SFRs in each bin. The purple band represents the epoch of reionization at $z \sim 6$--14 or 12.9--13.5 Gyr ago (Fan et al.\ 2006).}
\end{figure}

\begin{figure}
\includegraphics[width=1.0\columnwidth]{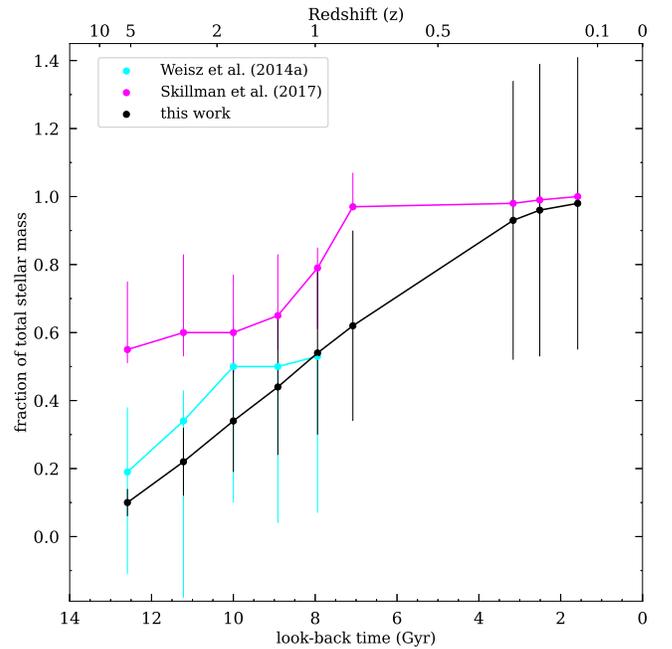}
\caption{Cumulative SFH as a function of look-back time (bottom axis) and redshift (top axis) within two half-light radii of And\,I for a metallicity of $Z=0.0007$, compared to the equivalent results from Weisz et al.\ (2014a) and Skillman et al.\ (2017).}
\end{figure}

Overall, And\,I appears similarly quenched to most of the low-mass satellites within the LG (Fillingham et al.\ 2018). Many processes have been described in the literature for the quenching of low-mass dwarf galaxies. Some of the main mechanisms are stellar feedback from supernov{\ae}, turbulence and galactic-scale winds as internal effects, and the epoch of cosmic reionization, tidal effects and ram pressure as external mechanisms (e.g., Weisz et al.\ 2014a; Fillingham et al.\ 2015). 
Fillingham et al.\ (2015) demonstrated that environmental processes are the main causes for quenching of low-mass satellites ($M_\star < 10^{5.5-8}$ M$_\odot$). They understood that, while the low-mass satellites in the Local Universe were quenched probably due to their proximity to their host galaxy, other low-mass dwarf galaxies can be star-forming and gas-rich. They defined quenched dwarf galaxies based on the observed H\,{\sc i} gas fraction if it is less than 10\% by mass. For And\,I, the atomic gas mass is estimated as $< 0.35\times10^6$ M$_\odot$ (Fillingham et al.\ 2015).

The delay time defines the time interval between infall (when a dwarf galaxy enters into the virial radius of its host galaxy) and quenching. It depends on how many environmental processes are effective in shutting down star formation in low-mass satellites. Overall, if the dwarf galaxy was quenched before its infall, then the host galaxy cannot have been responsible for quenching, otherwise, ram pressure stripping and tidal effects appear to be the main candidates for suppressing star formation in the few Gyr since infall.

Determining infall times requires simulation. Some have used the ELVIS suite of $N$-body simulations to determine the timescale on which satellites of the MW or M\,31 quench after falling into the virial radius ($\sim 300$ kpc; e.g., Fillingham et al.\ 2015, 2018; Wetzel et al.\ 2015b). Fillingham et al.\ (2019) determined the infall times and quenching timescales of MW satellites with Gaia proper motions along with ELVIS simulations. Although it was previously thought that a number of galaxies were quenched prior to becoming a satellite (e.g., Weisz et al.\ 2015), with more precise infall times, the role of the environment in quenching of star formation became more prominent (Fillingham et al.\ 2019).
For And\,I at a distance of 58 kpc from M\,31 (McConnachie 2012), we obtained a quenching time of $\sim 4$ Gyr ago. Assuming an infall time 5--8 Gyr ago estimated by Wetzel et al.\ (2015a) for LG dwarf galaxies with a range of stellar masses $10^{3-9}$ M$_\odot$, we can estimate a delay time of around 1 to 4 Gyr that is in good accordance with the literature (e.g., Wetzel et al.\ 2015b). In fact, our result clearly corroborates the environmental effects on galaxy quenching.

%
\subsection{Dark matter halo and SFH}

Due to being dominated by dark matter, dSph galaxies are noteworthy systems with total dynamical-light ratios of 10--1000 (e.g., Gilmore et al.\ 2007; Wolf et al.\ 2010). According to the standard $\Lambda$CDM model, the spatial distribution and dynamical properties of the dark matter halo strongly affect the galaxy formation and thus there is a strong correlation between the dark matter halo structure and star formation activity (e.g., Woo et al.\ 2008). 
Hayashi \& Chiba (2015) suggested a non-spherical density structure of dark matter haloes of dSph galaxies in the LG, using axisymmetric mass models. They assumed that the shapes of dark haloes in the dSphs are more elongated than those of $\Lambda$CDM subhaloes. Stellar feedback with gaseous outflow and tidal effects between the host galaxy and dark halo, together can change the shape of dark haloes. Hayashi \& Chiba (2015) investigated the dark halo properties using the time by which 70\% of the total stellar mass of the galaxy is formed as an indicator. dSphs which formed 70\% of their stellar mass early (rapid SFH) have a dense and concentrated dark matter halo, while others that formed it more recently have a low dark-matter density within the central region.

We recall that we obtain a time $\sim 6.20^{+1.57}_{-0.04}$ Gyr ago for And\,I to have created 70\% of its total stellar mass. Thus it looks like And\,I has a low dark-matter density close to its centre. In agreement with this, Kirby et al.\ (2020) calculated a ratio of total dynamical mass to stellar mass within the half-light radius of And\,I equal to $14^{+24}_{-9}$ that is lower than other dSph galaxies such as And\,V ($120^{+48}_{-34}$). (It should be clear that a galaxy without dark matter has a ratio $\sim1$).

\begin{table}
\caption[]{The fraction of total stellar mass formed prior to the specified epoch for And\,I based on the cumulative SFH.}
\begin{tabularx}{\linewidth}{cccc}
\hline \hline\\
$f_{\log t ({\rm yr})}$ \hspace{0.1 in} & Weisz et al.\  \hspace{0.1 in} & Skillman et al.\  \hspace{0.1 in} & this work \\
                  & (2014a)        &   (2017)          &           \\   
\hline\\
f$_{10.1}$ \hspace{0.1 in}& $0.19^{+0.3}_{-0.19}$  \hspace{0.1 in}& $0.55^{+0.04}_{-0.2}$ \hspace{0.1 in}& $0.10\pm0.04$ \vspace{0.05 in}\\
f$_{10.05}$\hspace{0.1 in}& $0.34^{+0.52}_{-0.09}$ \hspace{0.1 in}& $0.6^{+0.07}_{-0.23}$ \hspace{0.1 in}& $0.22\pm0.10$ \vspace{0.05 in}\\
f$_{10}$   \hspace{0.1 in}& $0.5^{+0.4}_{-0.14}$   \hspace{0.1 in}& $0.6^{+0.12}_{-0.17}$ \hspace{0.1 in}& $0.34\pm0.15$ \vspace{0.05 in}\\
f$_{9.95}$ \hspace{0.1 in}& $0.5^{+0.46}_{-0.0}$   \hspace{0.1 in}& $0.65^{+0.1}_{-0.18}$ \hspace{0.1 in}& $0.44\pm0.20$ \vspace{0.05 in}\\
f$_{9.9}$  \hspace{0.1 in}& $0.53^{+0.46}_{-0.0}$  \hspace{0.1 in}& $0.79^{+0.18}_{-0.06}$\hspace{0.1 in}& $0.54\pm0.24$ \vspace{0.05 in}\\
f$_{9.85}$ \hspace{0.1 in}&   .....                \hspace{0.1 in}& $0.97^{+0.01}_{-0.1}$ \hspace{0.1 in}& $0.62\pm0.28$ \vspace{0.05 in}\\
f$_{9.5}$  \hspace{0.1 in}&   .....                \hspace{0.1 in}& $0.98^{+0.01}_{-0.0}$ \hspace{0.1 in}& $0.93\pm0.41$ \vspace{0.05 in}\\
f$_{9.4}$  \hspace{0.1 in}&   .....                \hspace{0.1 in}& $0.99^{+0.0}_{-0.0}$  \hspace{0.1 in}& $0.96\pm0.43$ \vspace{0.05 in}\\
f$_{9.2}$  \hspace{0.1 in}&   .....                \hspace{0.1 in}& $1.00^{+0.0}_{-0.0}$  \hspace{0.1 in}& $0.98\pm0.43$ \vspace{0.05 in}\\
\hline\\
\end{tabularx}
\end{table}

%
\section{Conclusions}

The WFC at the INT was used to monitor the majority of dwarf galaxies in the LG including 43 dSph, six dIrr and six dTrans in the $i$-band filter with additional observations in the $V$ band over up to ten epochs. In this paper, we presented the methodology of estimating the SFH and all required relationships for the entire sample of monitored galaxies along with the SFH derived for And\,I dwarf galaxy as an example.

The photometric catalogue of Paper\,I was used to reconstruct the SFH of And\,I and estimate the total stellar mass of the galaxy. Using the numbers and luminosities of pulsating AGB stars and recent Padova stellar evolution models (Marigo et al.\ 2017), we obtained the SFR as a function of look-back time in this galaxy. Our conclusions about the SFH of And\,I can be summarised as follows:
\begin{itemize}
\item{Assuming a constant metallicity $Z=0.0007$, a total stellar mass of $5.9\pm2.7\times10^6$ M$_\odot$ is calculated within a half-light radius of And\,I ($3.2\pm0.3$ arcmin).}
\item{The SFH is characterised by a major epoch of formation peaking around 6.6 Gyr ago, assuming a constant metallicity of $Z=0.0007$, or 10.6 Gyr ago, assuming a lower metallicity of $Z=0.0002$.}
\item{A late epoch of star formation is seen in And\,I, peaking around 790 Myr ago, however the SFR is low ($6\pm3\times10^{-4}$ M$_\odot$ yr$^{-1}$). Feedback from dusty stellar winds at earlier times may be responsible for such late epoch of star formation in this quenched galaxy.}
\item{To examine how the SFH changes with the selection of our assumptions, different choices were tested including: using only reddened stars and high-amplitude candidates; and using different distance modulus, half light radius and metallicities. Overall, this does not affect the SFH of And\,I in a qualitative sense.}
\item{The cumulative SFH of And\,I shows that it took until $\sim 3.7^{+0.3}_{-1.0}$ Gyr ago to have formed 90\% of its total stellar mass and thus this galaxy had an extended SFH. Therefore, we have shown evidence for late quenching of this dSph satellite for the first time, possibly associated with late infall.}
\end{itemize}

\section*{Acknowledgments}

The observing time for this survey was primarily provided by the Iranian National Observatory (INO), complemented by UK-PATT allocation of time to programmes I/2016B/09 and I/2017B/04 (PI: J.\ van Loon). We thank the INO and the School of Astronomy (IPM) for the financial support of this project. We thank Alireza Molaeinezhad, Arash Danesh, Mojtaba Raouf, Ghassem Gozaliasl, James Bamber, Philip Short, Lucia Su\'arez-Andr\'es and Rosa Clavero for their help with the observations.


\appendix
\section{Supplementary material}
\label{app:app}

\begin{figure*}
\centering
\includegraphics[width=1.0\columnwidth]{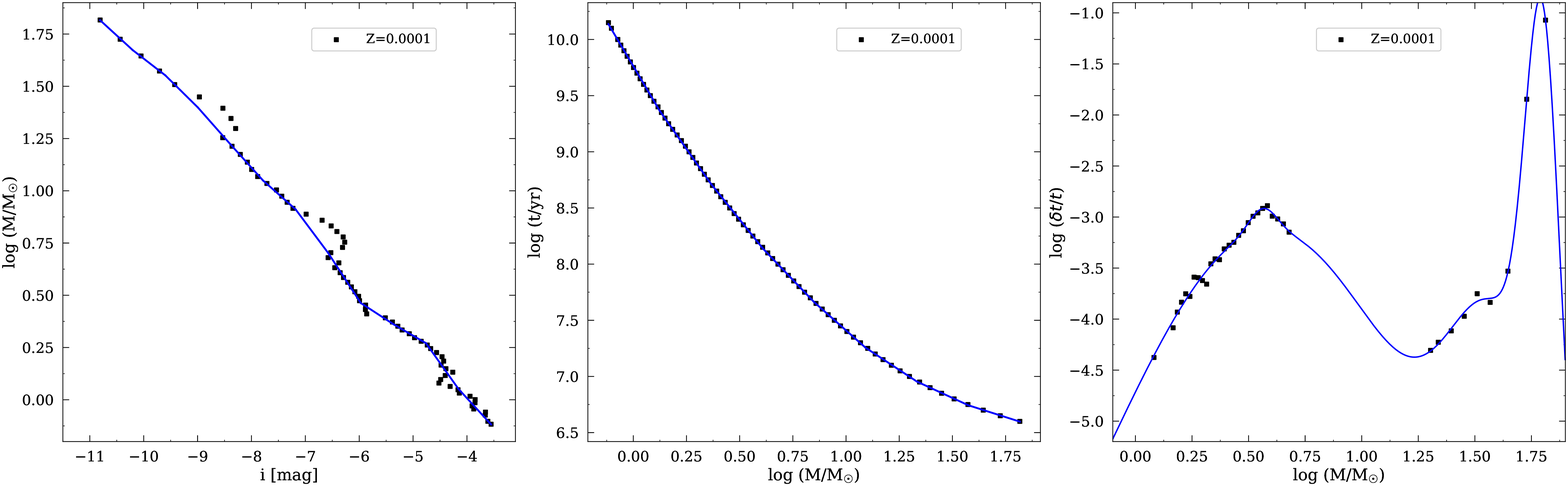}
\caption{Left panel: The mass--luminosity relation in the $i$ band for a metallicity of $Z=0.0001$. The linear spline fits, drawn in blue solid lines are obtained by minimisation of $\chi^2$ with the Iraf task {\sc gfit1d}. Middle panel: The mass--age relation for a metallicity of $Z=0.0001$ along with linear spline fits. Right panel: The mass--pulsation duration relation with the same metallicity, where the points show the ratio of pulsation duration to age {\it vs}.\ mass; the solid lines are multiple-Gaussian fits.}
\end{figure*}

\begin{figure*}
\centering
\includegraphics[width=1.0\columnwidth]{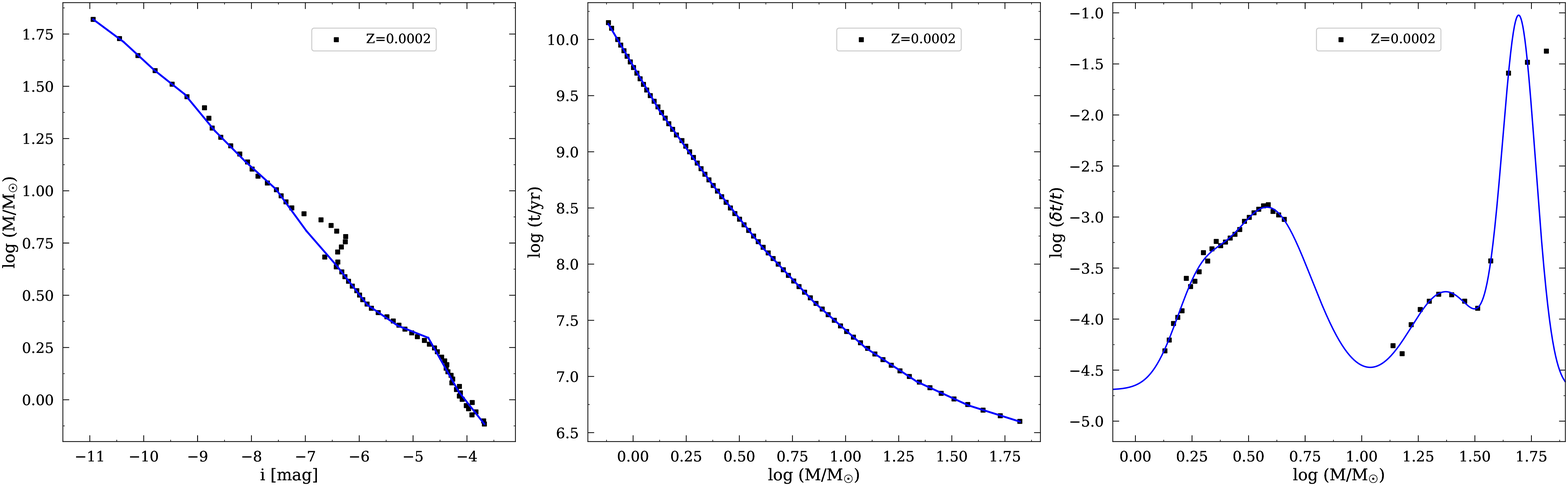}
\caption{Same as Figure A1 for a metallicity of $Z=0.0002$.}
\end{figure*}

\begin{figure*}
\centering
\includegraphics[width=1.0\columnwidth]{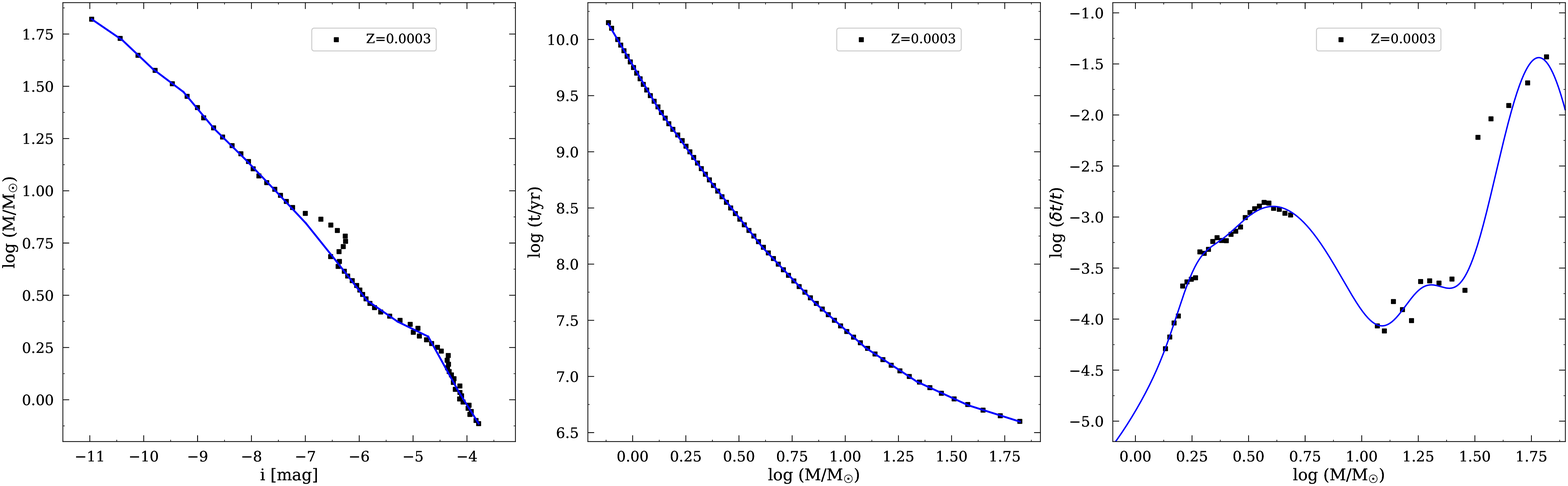}
\caption{Same as Figure A1 for a metallicity of $Z=0.0003$.}
\end{figure*}

\begin{figure*}
\centering
\includegraphics[width=1.0\columnwidth]{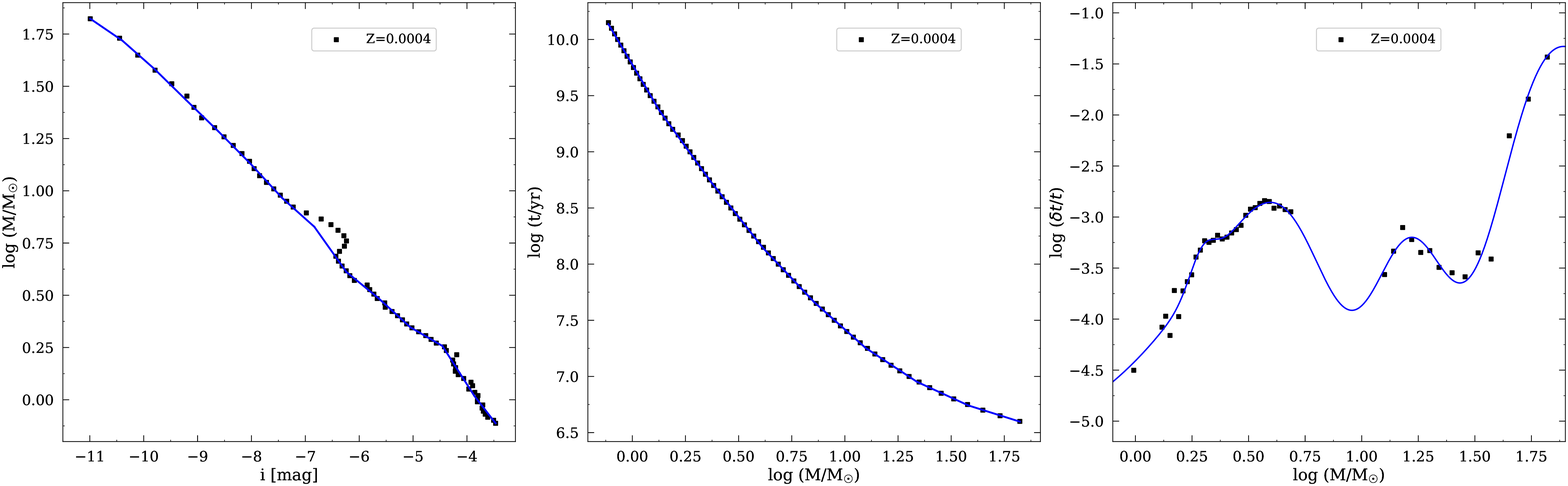}
\caption{Same as Figure A1 for a metallicity of $Z=0.0004$.}
\end{figure*}

\begin{figure*}
\centering
\includegraphics[width=1.0\columnwidth]{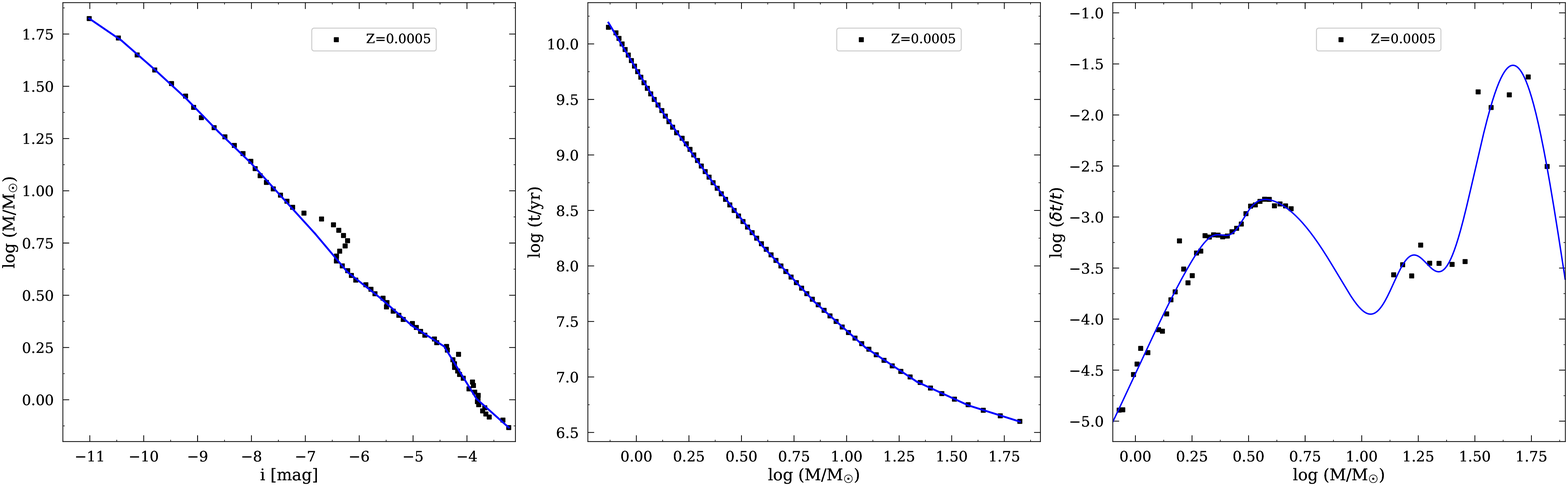}
\caption{Same as Figure A1 for a metallicity of $Z=0.0005$.}
\end{figure*}

\begin{figure*}
\centering
\includegraphics[width=1.0\columnwidth]{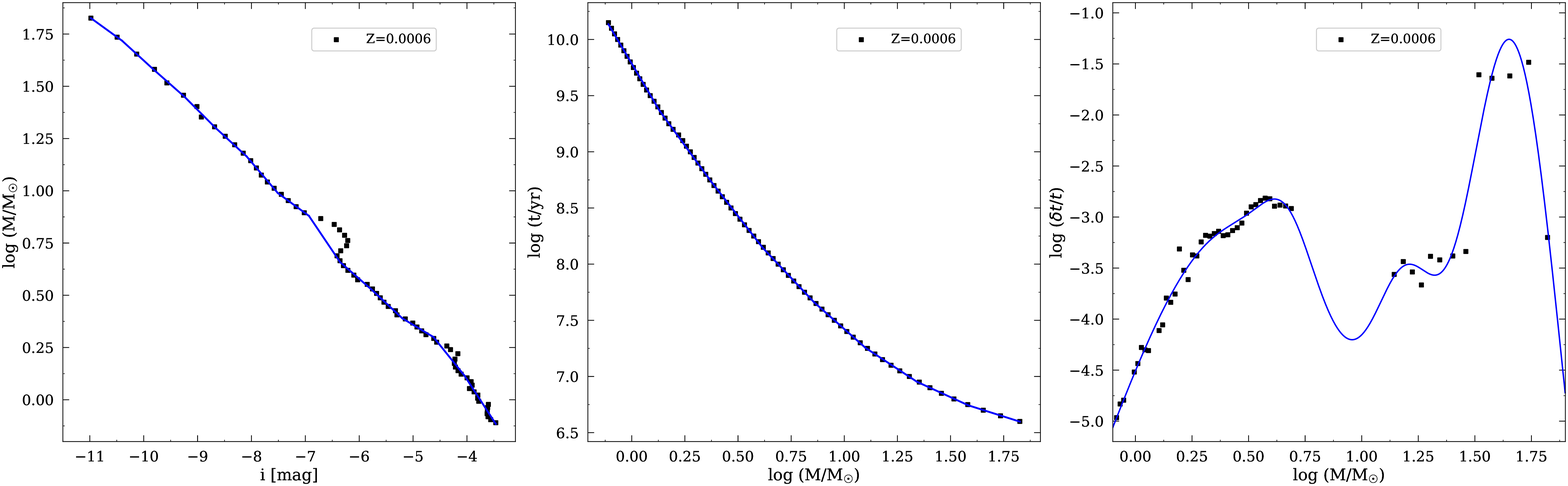}
\caption{Same as Figure A1 for a metallicity of $Z=0.0006$.}
\end{figure*}

\begin{figure*}
\centering
\includegraphics[width=1.0\columnwidth]{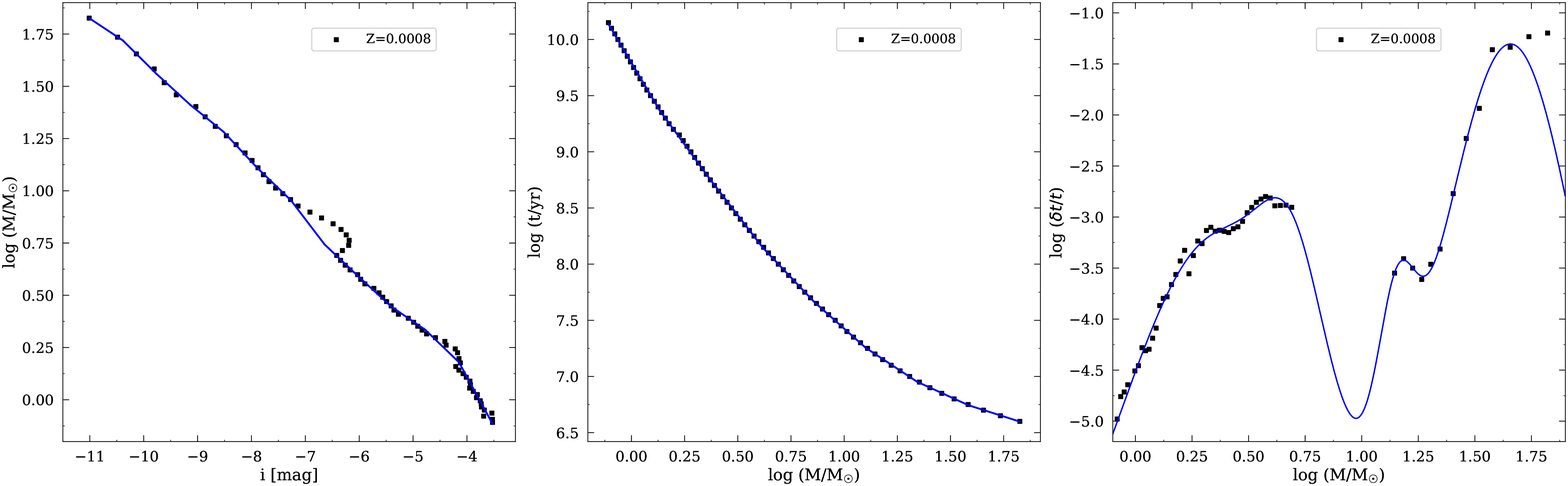}
\caption{Same as Figure A1 for a metallicity of $Z=0.0008$.}
\end{figure*}

\begin{figure*}
\centering
\includegraphics[width=1.0\columnwidth]{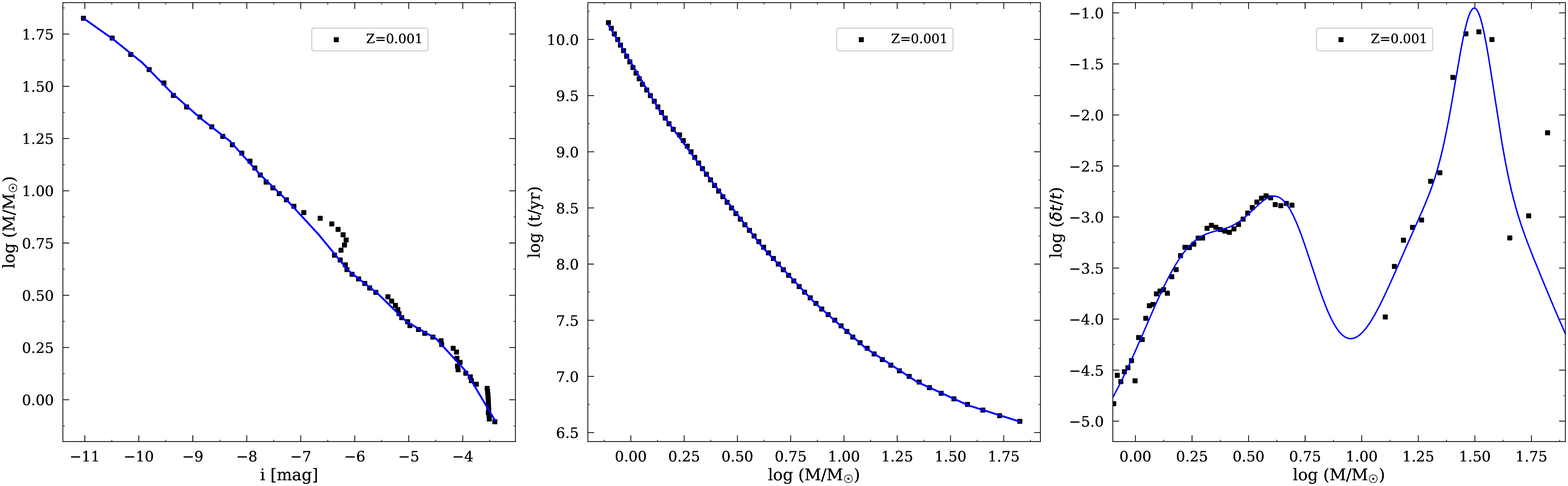}
\caption{Same as Figure A1 for a metallicity of $Z=0.001$.}
\end{figure*}

\begin{figure*}
\centering
\includegraphics[width=1.0\columnwidth]{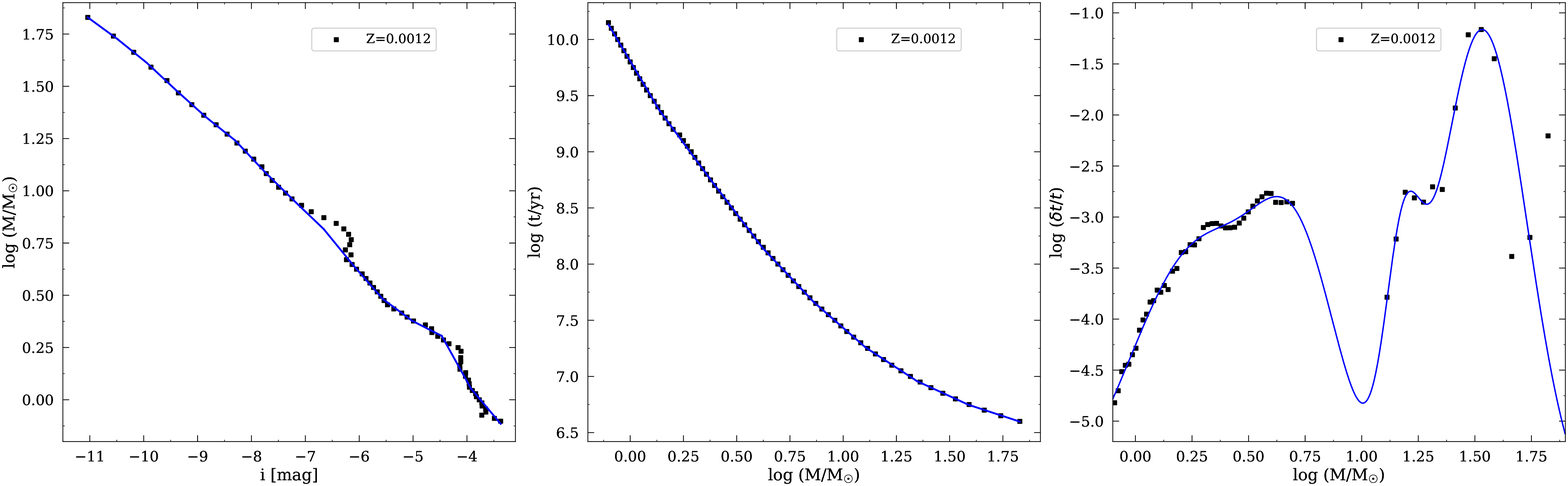}
\caption{Same as Figure A1 for a metallicity of $Z=0.0012$.}
\end{figure*}

\begin{figure*}
\centering
\includegraphics[width=1.0\columnwidth]{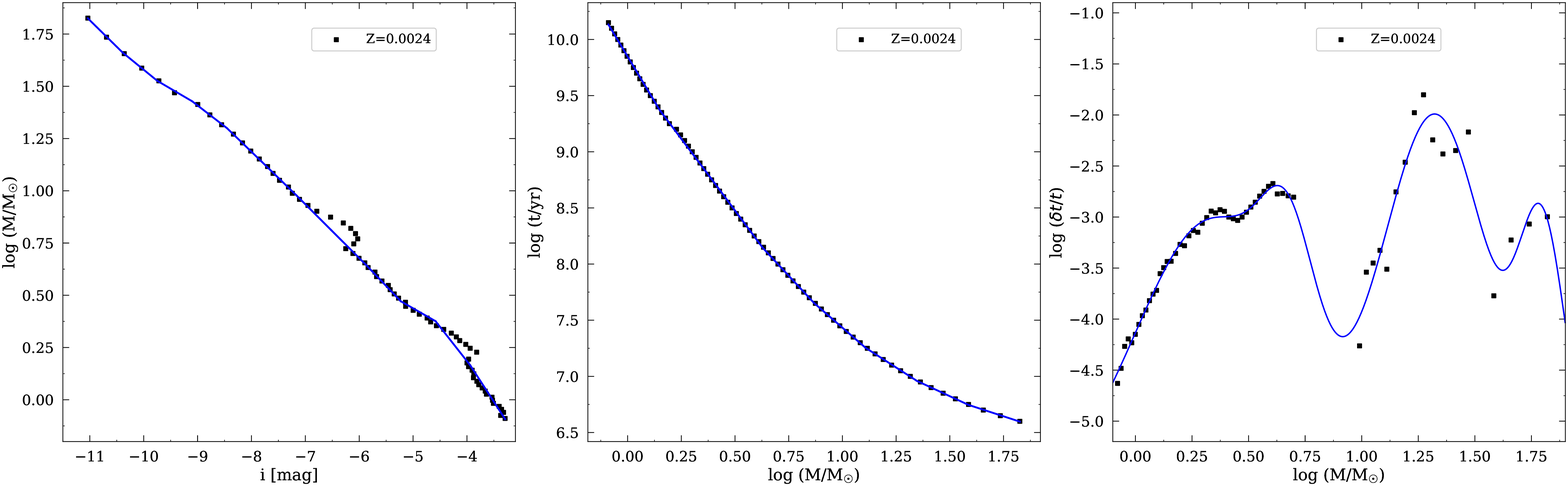}
\caption{Same as Figure A1 for a metallicity of $Z=0.0024$.}
\end{figure*}

\begin{figure*}
\centering
\includegraphics[width=1.0\columnwidth]{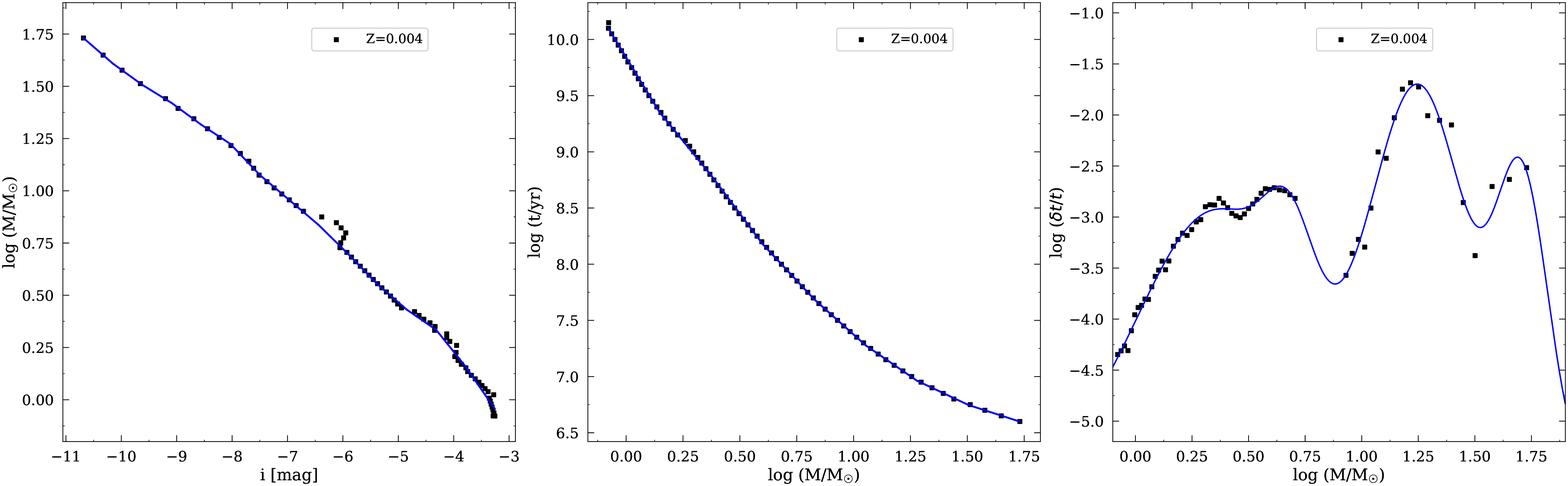}
\caption{Same as Figure A1 for a metallicity of $Z=0.004$.}
\end{figure*}

\begin{figure*}
\centering
\includegraphics[width=1.0\columnwidth]{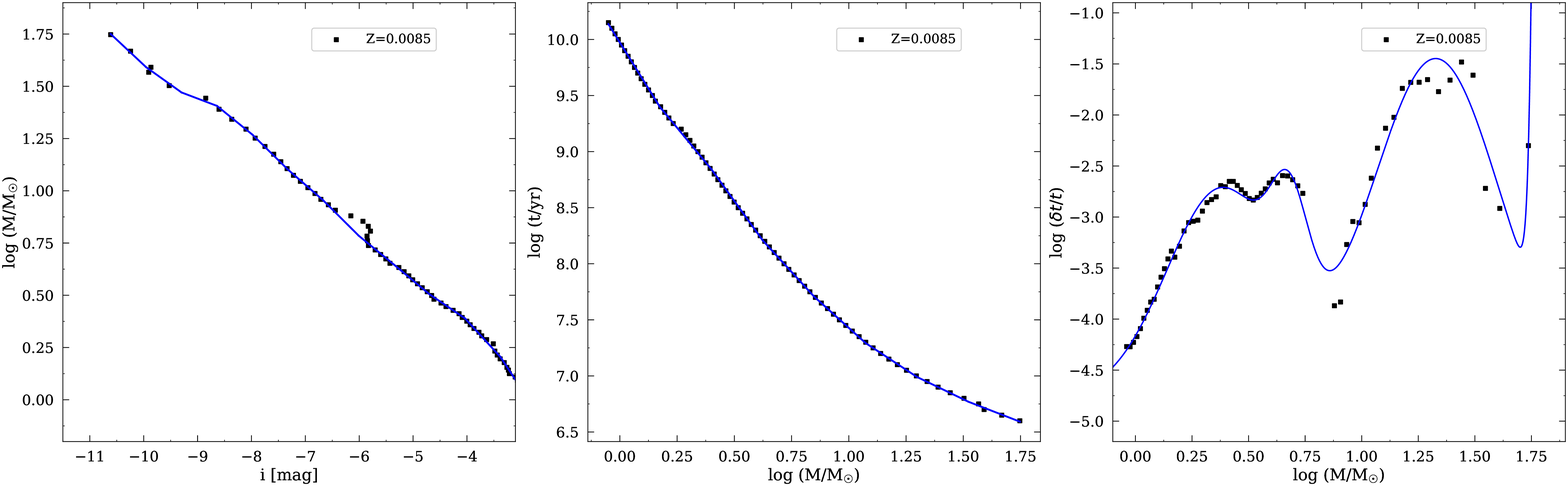}
\caption{Same as Figure A1 for a metallicity of $Z=0.0085$.}
\end{figure*}

\begin{table}
\caption{Fitting equations of the relation between birth mass and $i$-band magnitude, $\log M/M_\odot=ai+b$.}
\begin{tabular}{ccr}

\hline \hline \\
\hspace{0.05 in}$a$              & \hspace{0.08 in}$b$             & \hspace{0.08 in}validity range\vspace{0.05 in}\\
\hline\\
\multicolumn{3}{c}{$Z=0.0001$}\vspace{0.05 in} \\
\hline\\
\hspace{0.05 in}$ -0.236\pm0.076 $ & \hspace{0.1 in} $ -0.736\pm-0.752 $ & \hspace{0.1 in} $i\leq -10.204 $\vspace{0.04 in}\\
\hspace{0.05 in}$ -0.200\pm0.087 $ & \hspace{0.1 in} $ -0.364\pm-0.811 $ & \hspace{0.1 in} $ -10.204 <i\leq -9.599 $\vspace{0.04 in}\\
\hspace{0.05 in}$ -0.254\pm0.215 $ & \hspace{0.1 in} $ -0.889\pm-1.831 $ & \hspace{0.1 in} $ -9.599 <i\leq -8.995 $\vspace{0.04 in}\\
\hspace{0.05 in}$ -0.292\pm0.199 $ & \hspace{0.1 in} $ -1.225\pm-1.650 $ & \hspace{0.1 in} $ -8.995 <i\leq -8.390 $\vspace{0.04 in}\\
\hspace{0.05 in}$ -0.286\pm0.055 $ & \hspace{0.1 in} $ -1.176\pm-0.414 $ & \hspace{0.1 in} $ -8.390 <i\leq -7.785 $\vspace{0.04 in}\\
\hspace{0.05 in}$ -0.228\pm0.054 $ & \hspace{0.1 in} $ -0.729\pm-0.369 $ & \hspace{0.1 in} $ -7.785 <i\leq -7.180 $\vspace{0.04 in}\\
\hspace{0.05 in}$ -0.345\pm0.048 $ & \hspace{0.1 in} $ -1.569\pm-0.304 $ & \hspace{0.1 in} $ -7.180 <i\leq -6.576 $\vspace{0.04 in}\\
\hspace{0.05 in}$ -0.396\pm0.035 $ & \hspace{0.1 in} $ -1.899\pm-0.201 $ & \hspace{0.1 in} $ -6.576 <i\leq -5.971 $\vspace{0.04 in}\\
\hspace{0.05 in}$ -0.167\pm0.037 $ & \hspace{0.1 in} $ -0.537\pm-0.184 $ & \hspace{0.1 in} $ -5.971 <i\leq -5.366 $\vspace{0.04 in}\\
\hspace{0.05 in}$ -0.150\pm0.039 $ & \hspace{0.1 in} $ -0.444\pm-0.175 $ & \hspace{0.1 in} $ -5.366 <i\leq -4.761 $\vspace{0.04 in}\\
\hspace{0.05 in}$ -0.363\pm0.034 $ & \hspace{0.1 in} $ -1.456\pm-0.133 $ & \hspace{0.1 in} $ -4.761 <i\leq -4.157 $\vspace{0.04 in}\\
\hspace{0.05 in}$ -0.280\pm0.034 $ & \hspace{0.1 in} $ -1.114\pm-0.109 $ & \hspace{0.12 in} $i> -4.157 $\vspace{0.04 in}\\
\vspace{0.05 in}\\
\hline \hline \\
\multicolumn{3}{c}{$Z=0.0002$} \vspace{0.05 in} \\
\hline\\
\hspace{0.07 in}$ -0.191\pm0.057 $ & \hspace{0.12 in} $ -0.271\pm-0.580 $ & \hspace{0.12 in} $i\leq -10.373 $ \vspace{0.04 in}\\
\hspace{0.07 in}$ -0.237\pm0.057 $ & \hspace{0.12 in} $ -0.747\pm-0.542 $ & \hspace{0.12 in} $ -10.373 <i\leq -9.807 $\vspace{0.04 in}\\
\hspace{0.07 in}$ -0.205\pm0.080 $ & \hspace{0.12 in} $ -0.428\pm-0.712 $ & \hspace{0.12 in} $ -9.807 <i\leq -9.242 $\vspace{0.04 in}\\
\hspace{0.07 in}$ -0.312\pm0.075 $ & \hspace{0.12 in} $ -1.421\pm-0.637 $ & \hspace{0.12 in} $ -9.242 <i\leq -8.676 $\vspace{0.04 in}\\
\hspace{0.07 in}$ -0.259\pm0.040 $ & \hspace{0.12 in} $ -0.964\pm-0.318 $ & \hspace{0.12 in} $ -8.676 <i\leq -8.111 $\vspace{0.04 in}\\
\hspace{0.07 in}$ -0.233\pm0.037 $ & \hspace{0.12 in} $ -0.750\pm-0.270 $ & \hspace{0.12 in} $ -8.111 <i\leq -7.545 $\vspace{0.04 in}\\
\hspace{0.07 in}$ -0.353\pm0.063 $ & \hspace{0.12 in} $ -1.658\pm-0.418 $ & \hspace{0.12 in} $ -7.545 <i\leq -6.980 $\vspace{0.04 in}\\
\hspace{0.07 in}$ -0.296\pm0.059 $ & \hspace{0.12 in} $ -1.260\pm-0.370 $ & \hspace{0.12 in} $ -6.980 <i\leq -6.414 $\vspace{0.04 in}\\
\hspace{0.07 in}$ -0.329\pm0.029 $ & \hspace{0.12 in} $ -1.471\pm-0.163 $ & \hspace{0.12 in} $ -6.414 <i\leq -5.849 $\vspace{0.04 in}\\
\hspace{0.07 in}$ -0.176\pm0.031 $ & \hspace{0.12 in} $ -0.577\pm-0.155 $ & \hspace{0.12 in} $ -5.849 <i\leq -5.283 $\vspace{0.04 in}\\
\hspace{0.07 in}$ -0.102\pm0.031 $ & \hspace{0.12 in} $ -0.186\pm-0.138 $ & \hspace{0.12 in} $ -5.283 <i\leq -4.718 $\vspace{0.04 in}\\
\hspace{0.07 in}$ -0.457\pm0.024 $ & \hspace{0.12 in} $ -1.859\pm-0.096 $ & \hspace{0.12 in} $ -4.718 <i\leq -4.152 $\vspace{0.04 in}\\
\hspace{0.07 in}$ -0.273\pm0.028 $ & \hspace{0.12 in} $ -1.094\pm-0.091 $ & \hspace{0.12 in} $i> -4.152 $\vspace{0.05 in}\\
\hline \hline \\
\multicolumn{3}{c}{$Z=0.0003$} \vspace{0.05 in} \\
\hline\\
\hspace{0.05 in}$ -0.176\pm0.060 $ & \hspace{0.1 in} $ -0.109\pm-0.608 $ & \hspace{0.1 in} $i\leq -10.398 $\vspace{0.04 in}\\
\hspace{0.05 in}$ -0.244\pm0.058 $ & \hspace{0.1 in} $ -0.812\pm-0.557 $ & \hspace{0.1 in} $ -10.398 <i\leq -9.830 $\vspace{0.04 in}\\
\hspace{0.05 in}$ -0.194\pm0.055 $ & \hspace{0.1 in} $ -0.321\pm-0.491 $ & \hspace{0.1 in} $ -9.830 <i\leq -9.262 $\vspace{0.04 in}\\
\hspace{0.05 in}$ -0.309\pm0.048 $ & \hspace{0.1 in} $ -1.385\pm-0.402 $ & \hspace{0.1 in} $ -9.262 <i\leq -8.694 $\vspace{0.04 in}\\
\hspace{0.05 in}$ -0.253\pm0.041 $ & \hspace{0.1 in} $ -0.906\pm-0.323 $ & \hspace{0.1 in} $ -8.694 <i\leq -8.126 $\vspace{0.04 in}\\
\hspace{0.05 in}$ -0.269\pm0.041 $ & \hspace{0.1 in} $ -1.029\pm-0.301 $ & \hspace{0.1 in} $ -8.126 <i\leq -7.558 $\vspace{0.04 in}\\
\hspace{0.05 in}$ -0.275\pm0.104 $ & \hspace{0.1 in} $ -1.076\pm-0.681 $ & \hspace{0.1 in} $ -7.558 <i\leq -6.991 $\vspace{0.04 in}\\
\hspace{0.05 in}$ -0.322\pm0.099 $ & \hspace{0.1 in} $ -1.403\pm-0.628 $ & \hspace{0.1 in} $ -6.991 <i\leq -6.423 $\vspace{0.04 in}\\
\hspace{0.05 in}$ -0.330\pm0.032 $ & \hspace{0.1 in} $ -1.456\pm-0.179 $ & \hspace{0.1 in} $ -6.423 <i\leq -5.855 $\vspace{0.04 in}\\
\hspace{0.05 in}$ -0.178\pm0.037 $ & \hspace{0.1 in} $ -0.567\pm-0.183 $ & \hspace{0.1 in} $ -5.855 <i\leq -5.287 $\vspace{0.04 in}\\
\hspace{0.05 in}$ -0.126\pm0.044 $ & \hspace{0.1 in} $ -0.292\pm-0.197 $ & \hspace{0.1 in} $ -5.287 <i\leq -4.719 $\vspace{0.04 in}\\
\hspace{0.05 in}$ -0.467\pm0.033 $ & \hspace{0.1 in} $ -1.899\pm-0.131 $ & \hspace{0.1 in} $ -4.719 <i\leq -4.151 $\vspace{0.04 in}\\
\hspace{0.05 in}$ -0.266\pm0.031 $ & \hspace{0.1 in} $ -1.067\pm-0.100 $ & \hspace{0.1 in} $i> -4.151 $\vspace{0.05 in}\\
\hline \hline \\
\end{tabular}
\begin{tabular}{ccr}
\hline \hline \\
\hspace{0.05 in}$a$              & \hspace{0.08 in}$b$             & \hspace{0.08 in}validity range\vspace{0.05 in}\\
\hline\\
\multicolumn{3}{c}{$Z=0.0004$}\vspace{0.05 in} \\
\hline\\
\hspace{0.05 in}$ -0.176\pm0.052 $ & \hspace{0.1 in} $ -0.110\pm-0.521 $ & \hspace{0.1 in} $i\leq -10.398 $\vspace{0.04 in}\\
\hspace{0.05 in}$ -0.228\pm0.049 $ & \hspace{0.1 in} $ -0.648\pm-0.464 $ & \hspace{0.1 in} $ -10.398 <i\leq -9.804 $\vspace{0.04 in}\\
\hspace{0.05 in}$ -0.253\pm0.050 $ & \hspace{0.1 in} $ -0.893\pm-0.444 $ & \hspace{0.1 in} $ -9.804 <i\leq -9.210 $\vspace{0.04 in}\\
\hspace{0.05 in}$ -0.251\pm0.047 $ & \hspace{0.1 in} $ -0.875\pm-0.389 $ & \hspace{0.1 in} $ -9.210 <i\leq -8.615 $\vspace{0.04 in}\\
\hspace{0.05 in}$ -0.259\pm0.035 $ & \hspace{0.1 in} $ -0.950\pm-0.275 $ & \hspace{0.1 in} $ -8.615 <i\leq -8.021 $\vspace{0.04 in}\\
\hspace{0.05 in}$ -0.279\pm0.033 $ & \hspace{0.1 in} $ -1.107\pm-0.238 $ & \hspace{0.1 in} $ -8.021 <i\leq -7.427 $\vspace{0.04 in}\\
\hspace{0.05 in}$ -0.231\pm0.120 $ & \hspace{0.1 in} $ -0.748\pm-0.758 $ & \hspace{0.1 in} $ -7.427 <i\leq -6.833 $\vspace{0.04 in}\\
\hspace{0.05 in}$ -0.363\pm0.118 $ & \hspace{0.1 in} $ -1.655\pm-0.725 $ & \hspace{0.1 in} $ -6.833 <i\leq -6.239 $\vspace{0.04 in}\\
\hspace{0.05 in}$ -0.211\pm0.029 $ & \hspace{0.1 in} $ -0.704\pm-0.156 $ & \hspace{0.1 in} $ -6.239 <i\leq -5.645 $\vspace{0.04 in}\\
\hspace{0.05 in}$ -0.235\pm0.027 $ & \hspace{0.1 in} $ -0.839\pm-0.129 $ & \hspace{0.1 in} $ -5.645 <i\leq -5.050 $\vspace{0.04 in}\\
\hspace{0.05 in}$ -0.150\pm0.027 $ & \hspace{0.1 in} $ -0.413\pm-0.111 $ & \hspace{0.1 in} $ -5.050 <i\leq -4.456 $\vspace{0.04 in}\\
\hspace{0.05 in}$ -0.403\pm0.022 $ & \hspace{0.1 in} $ -1.537\pm-0.079 $ & \hspace{0.1 in} $ -4.456 <i\leq -3.862 $\vspace{0.04 in}\\
\hspace{0.05 in}$ -0.220\pm0.029 $ & \hspace{0.1 in} $ -0.832\pm-0.082 $ & \hspace{0.1 in} $i> -3.862 $\vspace{0.05 in}\\
\hline \hline \\
\multicolumn{3}{c}{$Z=0.0005$}\vspace{0.05 in} \\
\hline\\
\hspace{0.05 in}$ -0.176\pm0.054 $ & \hspace{0.1 in} $ -0.113\pm-0.548 $ & \hspace{0.1 in} $i\leq -10.412 $ \vspace{0.04 in}\\
\hspace{0.05 in}$ -0.225\pm0.051 $ & \hspace{0.1 in} $ -0.625\pm-0.487 $ & \hspace{0.1 in} $ -10.412 <i\leq -9.813 $ \vspace{0.04 in}\\
\hspace{0.05 in}$ -0.237\pm0.044 $ & \hspace{0.1 in} $ -0.745\pm-0.392 $ & \hspace{0.1 in} $ -9.813 <i\leq -9.214 $ \vspace{0.04 in}\\
\hspace{0.05 in}$ -0.265\pm0.040 $ & \hspace{0.1 in} $ -1.004\pm-0.336 $ & \hspace{0.1 in} $ -9.214 <i\leq -8.615 $ \vspace{0.04 in}\\
\hspace{0.05 in}$ -0.248\pm0.037 $ & \hspace{0.1 in} $ -0.853\pm-0.284 $ & \hspace{0.1 in} $ -8.615 <i\leq -8.016 $ \vspace{0.04 in}\\
\hspace{0.05 in}$ -0.284\pm0.034 $ & \hspace{0.1 in} $ -1.140\pm-0.242 $ & \hspace{0.1 in} $ -8.016 <i\leq -7.417 $ \vspace{0.04 in}\\
\hspace{0.05 in}$ -0.284\pm0.084 $ & \hspace{0.1 in} $ -1.138\pm-0.531 $ & \hspace{0.1 in} $ -7.417 <i\leq -6.818 $ \vspace{0.04 in}\\
\hspace{0.05 in}$ -0.306\pm0.081 $ & \hspace{0.1 in} $ -1.291\pm-0.497 $ & \hspace{0.1 in} $ -6.818 <i\leq -6.219 $ \vspace{0.04 in}\\
\hspace{0.05 in}$ -0.204\pm0.030 $ & \hspace{0.1 in} $ -0.655\pm-0.161 $ & \hspace{0.1 in} $ -6.219 <i\leq -5.620 $ \vspace{0.04 in}\\
\hspace{0.05 in}$ -0.225\pm0.029 $ & \hspace{0.1 in} $ -0.776\pm-0.136 $ & \hspace{0.1 in} $ -5.620 <i\leq -5.021 $ \vspace{0.04 in}\\
\hspace{0.05 in}$ -0.166\pm0.027 $ & \hspace{0.1 in} $ -0.480\pm-0.113 $ & \hspace{0.1 in} $ -5.021 <i\leq -4.422 $ \vspace{0.04 in}\\
\hspace{0.05 in}$ -0.418\pm0.022 $ & \hspace{0.1 in} $ -1.593\pm-0.079 $ & \hspace{0.1 in} $ -4.422 <i\leq -3.823 $ \vspace{0.04 in}\\
\hspace{0.05 in}$ -0.230\pm0.029 $ & \hspace{0.1 in} $ -0.873\pm-0.082 $ & \hspace{0.1 in} $i> -3.823 $ \vspace{0.05 in}\\
\hline \hline \\
\multicolumn{3}{c}{$Z=0.0006$}\vspace{0.05 in} \\
\hline\\
\hspace{0.05 in}$ -0.187\pm0.048 $ & \hspace{0.1 in} $ -0.225\pm-0.481 $ & \hspace{0.1 in} $i\leq -10.402 $ \vspace{0.04 in}\\
\hspace{0.05 in}$ -0.236\pm0.045 $ & \hspace{0.1 in} $ -0.734\pm-0.426 $ & \hspace{0.1 in} $ -10.402 <i\leq -9.824 $ \vspace{0.04 in}\\
\hspace{0.05 in}$ -0.227\pm0.039 $ & \hspace{0.1 in} $ -0.649\pm-0.353 $ & \hspace{0.1 in} $ -9.824 <i\leq -9.246 $ \vspace{0.04 in}\\
\hspace{0.05 in}$ -0.258\pm0.037 $ & \hspace{0.1 in} $ -0.936\pm-0.309 $ & \hspace{0.1 in} $ -9.246 <i\leq -8.668 $ \vspace{0.04 in}\\
\hspace{0.05 in}$ -0.237\pm0.033 $ & \hspace{0.1 in} $ -0.752\pm-0.256 $ & \hspace{0.1 in} $ -8.668 <i\leq -8.090 $ \vspace{0.04 in}\\
\hspace{0.05 in}$ -0.303\pm0.031 $ & \hspace{0.1 in} $ -1.283\pm-0.220 $ & \hspace{0.1 in} $ -8.090 <i\leq -7.512 $ \vspace{0.04 in}\\
\hspace{0.05 in}$ -0.189\pm0.039 $ & \hspace{0.1 in} $ -0.433\pm-0.254 $ & \hspace{0.1 in} $ -7.512 <i\leq -6.933 $ \vspace{0.04 in}\\
\hspace{0.05 in}$ -0.383\pm0.036 $ & \hspace{0.1 in} $ -1.774\pm-0.220 $ & \hspace{0.1 in} $ -6.933 <i\leq -6.355 $ \vspace{0.04 in}\\
\hspace{0.05 in}$ -0.225\pm0.025 $ & \hspace{0.1 in} $ -0.771\pm-0.138 $ & \hspace{0.1 in} $ -6.355 <i\leq -5.777 $ \vspace{0.04 in}\\
\hspace{0.05 in}$ -0.240\pm0.026 $ & \hspace{0.1 in} $ -0.860\pm-0.128 $ & \hspace{0.1 in} $ -5.777 <i\leq -5.199 $ \vspace{0.04 in}\\
\hspace{0.05 in}$ -0.155\pm0.027 $ & \hspace{0.1 in} $ -0.417\pm-0.116 $ & \hspace{0.1 in} $ -5.199 <i\leq -4.621 $ \vspace{0.04 in}\\
\hspace{0.05 in}$ -0.321\pm0.023 $ & \hspace{0.1 in} $ -1.184\pm-0.087 $ & \hspace{0.1 in} $ -4.621 <i\leq -4.043 $ \vspace{0.04 in}\\
\hspace{0.05 in}$ -0.398\pm0.021 $ & \hspace{0.1 in} $ -1.494\pm-0.066 $ & \hspace{0.1 in} $i> -4.043 $ \vspace{0.05 in}\\
\hline\hline\\
\end{tabular}
\end{table}

\begin{table}
\begin{tabular}{ccr}
\\
\hline \hline \\
\hspace{0.05 in}$a$              & \hspace{0.08 in}$b$             & \hspace{0.08 in}validity range\vspace{0.05 in}\\
\hline\\
\multicolumn{3}{c}{$Z=0.0008$}\vspace{0.05 in} \\
\hline\\
\hspace{0.05 in}$ -0.171\pm0.059 $ & \hspace{0.1 in} $ -0.057\pm-0.599 $ & \hspace{0.1 in} $i\leq -10.387 $\vspace{0.04 in}\\
\hspace{0.05 in}$ -0.256\pm0.054 $ & \hspace{0.1 in} $ -0.939\pm-0.508 $ & \hspace{0.1 in} $ -10.387 <i\leq -9.763 $\vspace{0.04 in}\\
\hspace{0.05 in}$ -0.237\pm0.051 $ & \hspace{0.1 in} $ -0.750\pm-0.450 $ & \hspace{0.1 in} $ -9.763 <i\leq -9.139 $\vspace{0.04 in}\\
\hspace{0.05 in}$ -0.209\pm0.049 $ & \hspace{0.1 in} $ -0.498\pm-0.407 $ & \hspace{0.1 in} $ -9.139 <i\leq -8.516 $\vspace{0.04 in}\\
\hspace{0.05 in}$ -0.276\pm0.039 $ & \hspace{0.1 in} $ -1.070\pm-0.298 $ & \hspace{0.1 in} $ -8.516 <i\leq -7.892 $\vspace{0.04 in}\\
\hspace{0.05 in}$ -0.248\pm0.036 $ & \hspace{0.1 in} $ -0.849\pm-0.252 $ & \hspace{0.1 in} $ -7.892 <i\leq -7.268 $\vspace{0.04 in}\\
\hspace{0.05 in}$ -0.338\pm0.055 $ & \hspace{0.1 in} $ -1.502\pm-0.341 $ & \hspace{0.1 in} $ -7.268 <i\leq -6.644 $\vspace{0.04 in}\\
\hspace{0.05 in}$ -0.244\pm0.052 $ & \hspace{0.1 in} $ -0.881\pm-0.304 $ & \hspace{0.1 in} $ -6.644 <i\leq -6.020 $\vspace{0.04 in}\\
\hspace{0.05 in}$ -0.235\pm0.031 $ & \hspace{0.1 in} $ -0.825\pm-0.159 $ & \hspace{0.1 in} $ -6.020 <i\leq -5.396 $\vspace{0.04 in}\\
\hspace{0.05 in}$ -0.176\pm0.030 $ & \hspace{0.1 in} $ -0.507\pm-0.134 $ & \hspace{0.1 in} $ -5.396 <i\leq -4.773 $\vspace{0.04 in}\\
\hspace{0.05 in}$ -0.248\pm0.026 $ & \hspace{0.1 in} $ -0.849\pm-0.101 $ & \hspace{0.1 in} $ -4.773 <i\leq -4.149 $\vspace{0.04 in}\\
\hspace{0.05 in}$ -0.472\pm0.022 $ & \hspace{0.1 in} $ -1.780\pm-0.072 $ & \hspace{0.1 in} $i> -4.149 $\vspace{0.04 in}\\
\hline \hline \\
\multicolumn{3}{c}{$Z=0.001$}\vspace{0.05 in} \\
\hline\\
\hspace{0.05 in}$ -0.183\pm0.055 $ & \hspace{0.1 in} $ -0.187\pm-0.557 $ & \hspace{0.1 in} $i\leq -10.481 $\vspace{0.04 in}\\
\hspace{0.05 in}$ -0.208\pm0.056 $ & \hspace{0.1 in} $ -0.452\pm-0.542 $ & \hspace{0.1 in} $ -10.481 <i\leq -9.936 $\vspace{0.04 in}\\
\hspace{0.05 in}$ -0.265\pm0.052 $ & \hspace{0.1 in} $ -1.016\pm-0.473 $ & \hspace{0.1 in} $ -9.936 <i\leq -9.392 $\vspace{0.04 in}\\
\hspace{0.05 in}$ -0.227\pm0.045 $ & \hspace{0.1 in} $ -0.668\pm-0.386 $ & \hspace{0.1 in} $ -9.392 <i\leq -8.848 $\vspace{0.04 in}\\
\hspace{0.05 in}$ -0.201\pm0.043 $ & \hspace{0.1 in} $ -0.435\pm-0.346 $ & \hspace{0.1 in} $ -8.848 <i\leq -8.304 $\vspace{0.04 in}\\
\hspace{0.05 in}$ -0.288\pm0.037 $ & \hspace{0.1 in} $ -1.154\pm-0.277 $ & \hspace{0.1 in} $ -8.304 <i\leq -7.759 $\vspace{0.04 in}\\
\hspace{0.05 in}$ -0.249\pm0.036 $ & \hspace{0.1 in} $ -0.851\pm-0.250 $ & \hspace{0.1 in} $ -7.759 <i\leq -7.215 $\vspace{0.04 in}\\
\hspace{0.05 in}$ -0.280\pm0.079 $ & \hspace{0.1 in} $ -1.079\pm-0.492 $ & \hspace{0.1 in} $ -7.215 <i\leq -6.671 $\vspace{0.04 in}\\
\hspace{0.05 in}$ -0.314\pm0.076 $ & \hspace{0.1 in} $ -1.305\pm-0.455 $ & \hspace{0.1 in} $ -6.671 <i\leq -6.126 $\vspace{0.04 in}\\
\hspace{0.05 in}$ -0.203\pm0.037 $ & \hspace{0.1 in} $ -0.621\pm-0.198 $ & \hspace{0.1 in} $ -6.126 <i\leq -5.582 $\vspace{0.04 in}\\
\hspace{0.05 in}$ -0.251\pm0.035 $ & \hspace{0.1 in} $ -0.891\pm-0.168 $ & \hspace{0.1 in} $ -5.582 <i\leq -5.038 $\vspace{0.04 in}\\
\hspace{0.05 in}$ -0.149\pm0.032 $ & \hspace{0.1 in} $ -0.377\pm-0.132 $ & \hspace{0.1 in} $ -5.038 <i\leq -4.494 $\vspace{0.04 in}\\
\hspace{0.05 in}$ -0.282\pm0.030 $ & \hspace{0.1 in} $ -0.976\pm-0.112 $ & \hspace{0.1 in} $ -4.494 <i\leq -3.949 $\vspace{0.04 in}\\
\hspace{0.05 in}$ -0.435\pm0.025 $ & \hspace{0.1 in} $ -1.578\pm-0.080 $ & \hspace{0.1 in} $i> -3.949 $\vspace{0.05 in}\\
\hline \hline \\
\multicolumn{3}{c}{$Z=0.0012$}\vspace{0.05 in} \\
\hline\\
\hspace{0.05 in}$ -0.188\pm0.052 $ & \hspace{0.1 in} $ -0.250\pm-0.536 $ & \hspace{0.1 in} $i\leq -10.491 $\vspace{0.04 in}\\
\hspace{0.05 in}$ -0.209\pm0.052 $ & \hspace{0.1 in} $ -0.463\pm-0.508 $ & \hspace{0.1 in} $ -10.491 <i\leq -9.943 $\vspace{0.04 in}\\
\hspace{0.05 in}$ -0.241\pm0.047 $ & \hspace{0.1 in} $ -0.782\pm-0.430 $ & \hspace{0.1 in} $ -9.943 <i\leq -9.396 $\vspace{0.04 in}\\
\hspace{0.05 in}$ -0.232\pm0.043 $ & \hspace{0.1 in} $ -0.702\pm-0.366 $ & \hspace{0.1 in} $ -9.396 <i\leq -8.848 $\vspace{0.04 in}\\
\hspace{0.05 in}$ -0.206\pm0.040 $ & \hspace{0.1 in} $ -0.469\pm-0.318 $ & \hspace{0.1 in} $ -8.848 <i\leq -8.301 $\vspace{0.04 in}\\
\hspace{0.05 in}$ -0.274\pm0.034 $ & \hspace{0.1 in} $ -1.032\pm-0.256 $ & \hspace{0.1 in} $ -8.301 <i\leq -7.753 $\vspace{0.04 in}\\
\hspace{0.05 in}$ -0.254\pm0.035 $ & \hspace{0.1 in} $ -0.878\pm-0.244 $ & \hspace{0.1 in} $ -7.753 <i\leq -7.206 $\vspace{0.04 in}\\
\hspace{0.05 in}$ -0.247\pm0.108 $ & \hspace{0.1 in} $ -0.825\pm-0.670 $ & \hspace{0.1 in} $ -7.206 <i\leq -6.659 $\vspace{0.04 in}\\
\hspace{0.05 in}$ -0.316\pm0.104 $ & \hspace{0.1 in} $ -1.286\pm-0.625 $ & \hspace{0.1 in} $ -6.659 <i\leq -6.111 $\vspace{0.04 in}\\
\hspace{0.05 in}$ -0.295\pm0.028 $ & \hspace{0.1 in} $ -1.157\pm-0.151 $ & \hspace{0.1 in} $ -6.111 <i\leq -5.564 $\vspace{0.04 in}\\
\hspace{0.05 in}$ -0.191\pm0.029 $ & \hspace{0.1 in} $ -0.582\pm-0.138 $ & \hspace{0.1 in} $ -5.564 <i\leq -5.016 $\vspace{0.04 in}\\
\hspace{0.05 in}$ -0.131\pm0.033 $ & \hspace{0.1 in} $ -0.281\pm-0.139 $ & \hspace{0.1 in} $ -5.016 <i\leq -4.469 $\vspace{0.04 in}\\
\hspace{0.05 in}$ -0.463\pm0.026 $ & \hspace{0.1 in} $ -1.764\pm-0.096 $ & \hspace{0.1 in} $ -4.469 <i\leq -3.921 $\vspace{0.04 in}\\
\hspace{0.05 in}$ -0.306\pm0.026 $ & \hspace{0.1 in} $ -1.146\pm-0.079 $ & \hspace{0.1 in} $i> -3.921 $\vspace{0.05 in}\\
\hline \hline \\
\end{tabular}
\begin{tabular}{ccr}
\\
\hline \hline \\
\hspace{0.05 in}$a$              & \hspace{0.08 in}$b$             & \hspace{0.08 in}validity range\vspace{0.05 in}\\
\hline\\
\multicolumn{3}{c}{$Z=0.0024$}\vspace{0.05 in} \\
\hline\\
\hspace{0.05 in}$ -0.254\pm0.057 $ & \hspace{0.1 in} $ -0.973\pm-0.571 $ & \hspace{0.1 in} $i\leq -10.392 $\vspace{0.04 in}\\
\hspace{0.05 in}$ -0.210\pm0.055 $ & \hspace{0.1 in} $ -0.523\pm-0.516 $ & \hspace{0.1 in} $ -10.392 <i\leq -9.746 $\vspace{0.04 in}\\
\hspace{0.05 in}$ -0.151\pm0.058 $ & \hspace{0.1 in} $ 0.053\pm-0.510 $ & \hspace{0.1 in} $ -9.746 <i\leq -9.101 $\vspace{0.04 in}\\
\hspace{0.05 in}$ -0.199\pm0.055 $ & \hspace{0.1 in} $ -0.383\pm-0.452 $ & \hspace{0.1 in} $ -9.101 <i\leq -8.455 $\vspace{0.04 in}\\
\hspace{0.05 in}$ -0.249\pm0.044 $ & \hspace{0.1 in} $ -0.804\pm-0.331 $ & \hspace{0.1 in} $ -8.455 <i\leq -7.810 $\vspace{0.04 in}\\
\hspace{0.05 in}$ -0.250\pm0.039 $ & \hspace{0.1 in} $ -0.813\pm-0.269 $ & \hspace{0.1 in} $ -7.810 <i\leq -7.165 $\vspace{0.04 in}\\
\hspace{0.05 in}$ -0.258\pm0.055 $ & \hspace{0.1 in} $ -0.868\pm-0.336 $ & \hspace{0.1 in} $ -7.165 <i\leq -6.519 $\vspace{0.04 in}\\
\hspace{0.05 in}$ -0.257\pm0.052 $ & \hspace{0.1 in} $ -0.866\pm-0.294 $ & \hspace{0.1 in} $ -6.519 <i\leq -5.874 $\vspace{0.04 in}\\
\hspace{0.05 in}$ -0.270\pm0.032 $ & \hspace{0.1 in} $ -0.938\pm-0.155 $ & \hspace{0.1 in} $ -5.874 <i\leq -5.228 $\vspace{0.04 in}\\
\hspace{0.05 in}$ -0.148\pm0.034 $ & \hspace{0.1 in} $ -0.302\pm-0.145 $ & \hspace{0.1 in} $ -5.228 <i\leq -4.583 $\vspace{0.04 in}\\
\hspace{0.05 in}$ -0.328\pm0.030 $ & \hspace{0.1 in} $ -1.127\pm-0.111 $ & \hspace{0.1 in} $ -4.583 <i\leq -3.937 $\vspace{0.04 in}\\
\hspace{0.05 in}$ -0.398\pm0.024 $ & \hspace{0.12 in} $ -1.402\pm-0.071 $ & \hspace{0.12 in} $i> -3.937 $\vspace{0.04 in}\\
\hline \hline \\
\multicolumn{3}{c}{$Z=0.004$}\vspace{0.05 in} \\
\hline\\
\hspace{0.05 in}$ -0.230\pm0.068 $ & \hspace{0.1 in} $ -0.729\pm-0.669 $ & \hspace{0.1 in} $i\leq -10.153 $\vspace{0.04 in}\\
\hspace{0.05 in}$ -0.192\pm0.070 $ & \hspace{0.1 in} $ -0.336\pm-0.656 $ & \hspace{0.1 in} $ -10.153 <i\leq -9.622 $\vspace{0.04 in}\\
\hspace{0.05 in}$ -0.164\pm0.064 $ & \hspace{0.1 in} $ -0.068\pm-0.565 $ & \hspace{0.1 in} $ -9.622 <i\leq -9.092 $\vspace{0.04 in}\\
\hspace{0.05 in}$ -0.195\pm0.060 $ & \hspace{0.1 in} $ -0.351\pm-0.497 $ & \hspace{0.1 in} $ -9.092 <i\leq -8.561 $\vspace{0.04 in}\\
\hspace{0.05 in}$ -0.175\pm0.055 $ & \hspace{0.1 in} $ -0.182\pm-0.425 $ & \hspace{0.1 in} $ -8.561 <i\leq -8.031 $\vspace{0.04 in}\\
\hspace{0.05 in}$ -0.279\pm0.045 $ & \hspace{0.1 in} $ -1.020\pm-0.326 $ & \hspace{0.1 in} $ -8.031 <i\leq -7.500 $\vspace{0.04 in}\\
\hspace{0.05 in}$ -0.225\pm0.043 $ & \hspace{0.1 in} $ -0.614\pm-0.288 $ & \hspace{0.1 in} $ -7.500 <i\leq -6.970 $\vspace{0.04 in}\\
\hspace{0.05 in}$ -0.226\pm0.085 $ & \hspace{0.1 in} $ -0.616\pm-0.513 $ & \hspace{0.1 in} $ -6.970 <i\leq -6.440 $\vspace{0.04 in}\\
\hspace{0.05 in}$ -0.260\pm0.081 $ & \hspace{0.1 in} $ -0.839\pm-0.468 $ & \hspace{0.1 in} $ -6.440 <i\leq -5.909 $\vspace{0.04 in}\\
\hspace{0.05 in}$ -0.274\pm0.036 $ & \hspace{0.1 in} $ -0.921\pm-0.184 $ & \hspace{0.1 in} $ -5.909 <i\leq -5.379 $\vspace{0.04 in}\\
\hspace{0.05 in}$ -0.229\pm0.039 $ & \hspace{0.1 in} $ -0.681\pm-0.176 $ & \hspace{0.1 in} $ -5.379 <i\leq -4.848 $\vspace{0.04 in}\\
\hspace{0.05 in}$ -0.182\pm0.048 $ & \hspace{0.1 in} $ -0.451\pm-0.195 $ & \hspace{0.1 in} $ -4.848 <i\leq -4.318 $\vspace{0.04 in}\\
\hspace{0.05 in}$ -0.331\pm0.041 $ & \hspace{0.1 in} $ -1.095\pm-0.148 $ & \hspace{0.1 in} $ -4.318 <i\leq -3.787 $\vspace{0.04 in}\\
\hspace{0.05 in}$ -0.381\pm0.026 $ & \hspace{0.1 in} $ -1.284\pm-0.077 $ & \hspace{0.1 in} $i> -3.787 $\vspace{0.05 in}\\
\hline \hline \\
\multicolumn{3}{c}{$Z=0.0085$}\vspace{0.05 in} \\
\hline\\
\vspace{0.05 in}\\
\hspace{0.05 in}$ -0.241\pm0.022 $ & \hspace{0.1 in} $ -0.810\pm-0.216 $ & \hspace{0.1 in} $i\leq -9.955 $\vspace{0.04 in}\\
\hspace{0.05 in}$ -0.184\pm0.030 $ & \hspace{0.1 in} $ -0.240\pm-0.266 $ & \hspace{0.1 in} $ -9.955 <i\leq -9.298 $\vspace{0.04 in}\\
\hspace{0.05 in}$ -0.098\pm0.030 $ & \hspace{0.1 in} $ 0.561\pm-0.256 $ & \hspace{0.1 in} $ -9.298 <i\leq -8.640 $\vspace{0.04 in}\\
\hspace{0.05 in}$ -0.209\pm0.020 $ & \hspace{0.1 in} $ -0.401\pm-0.152 $ & \hspace{0.1 in} $ -8.640 <i\leq -7.982 $\vspace{0.04 in}\\
\hspace{0.05 in}$ -0.254\pm0.017 $ & \hspace{0.1 in} $ -0.756\pm-0.118 $ & \hspace{0.1 in} $ -7.982 <i\leq -7.325 $\vspace{0.04 in}\\
\hspace{0.05 in}$ -0.225\pm0.015 $ & \hspace{0.1 in} $ -0.544\pm-0.098 $ & \hspace{0.1 in} $ -7.325 <i\leq -6.667 $\vspace{0.04 in}\\
\hspace{0.05 in}$ -0.257\pm0.019 $ & \hspace{0.1 in} $ -0.758\pm-0.108 $ & \hspace{0.1 in} $ -6.667 <i\leq -6.009 $\vspace{0.04 in}\\
\hspace{0.05 in}$ -0.212\pm0.019 $ & \hspace{0.1 in} $ -0.486\pm-0.096 $ & \hspace{0.1 in} $ -6.009 <i\leq -5.352 $\vspace{0.04 in}\\
\hspace{0.05 in}$ -0.215\pm0.014 $ & \hspace{0.1 in} $ -0.505\pm-0.061 $ & \hspace{0.1 in} $ -5.352 <i\leq -4.694 $\vspace{0.04 in}\\
\hspace{0.05 in}$ -0.175\pm0.012 $ & \hspace{0.1 in} $ -0.317\pm-0.046 $ & \hspace{0.1 in} $ -4.694 <i\leq -4.036 $\vspace{0.04 in}\\
\hspace{0.05 in}$ -0.280\pm0.011 $ & \hspace{0.1 in} $ -0.743\pm-0.034 $ & \hspace{0.1 in} $ -4.036 <i\leq -3.379 $\vspace{0.04 in}\\
\hspace{0.05 in}$ -0.392\pm0.011 $ & \hspace{0.12 in} $ -1.120\pm-0.026 $ & \hspace{0.12 in} $i> -3.379 $\vspace{0.05 in}\\
\vspace{0.05 in}\\
\hline\\
\end{tabular}
\label{tab:taba1}
\end{table}

\begin{table}
\caption[]{Fitting equations of the relation between age and birth mass, $\log t=a\log M+b$.}
\begin{tabular}{ccr}
\\
\hline \hline\\
\hspace{0.05 in}$a$              & \hspace{0.09 in}$b$             & \hspace{0.1 in}validity range\vspace{0.05 in}\\
\hline\\
\multicolumn{3}{c}{$Z=0.0001$} \vspace{0.05 in}\\
\hline\\
\hspace{0.07 in}$ -3.200\pm0.019 $ & \hspace{0.09 in} $ 9.762\pm0.004 $ & \hspace{0.1 in} $\log M\leq 0.125 $\vspace{0.04 in}\\
\hspace{0.07 in}$ -2.697\pm0.017 $ & \hspace{0.09 in} $ 9.699\pm0.008 $ & \hspace{0.1 in} $ 0.125 <\log M\leq 0.366 $\vspace{0.04 in}\\
\hspace{0.07 in}$ -2.359\pm0.018 $ & \hspace{0.09 in} $ 9.576\pm0.013 $ & \hspace{0.1 in} $ 0.366 <\log M\leq 0.608 $\vspace{0.04 in}\\
\hspace{0.07 in}$ -1.982\pm0.020 $ & \hspace{0.09 in} $ 9.347\pm0.019 $ & \hspace{0.1 in} $ 0.608 <\log M\leq 0.850 $\vspace{0.04 in}\\
\hspace{0.07 in}$ -1.678\pm0.022 $ & \hspace{0.09 in} $ 9.088\pm0.027 $ & \hspace{0.1 in} $ 0.850 <\log M\leq 1.092 $\vspace{0.04 in}\\
\hspace{0.07 in}$ -1.249\pm0.025 $ & \hspace{0.09 in} $ 8.619\pm0.037 $ & \hspace{0.1 in} $ 1.092 <\log M\leq 1.333 $\vspace{0.04 in}\\
\hspace{0.07 in}$ -0.872\pm0.029 $ & \hspace{0.09 in} $ 8.117\pm0.050 $ & \hspace{0.1 in} $ 1.333 <\log M\leq 1.575 $\vspace{0.04 in}\\
\hspace{0.07 in}$ -0.602\pm0.035 $ & \hspace{0.09 in} $ 7.691\pm0.069 $ & \hspace{0.1 in} $\log M> 1.575 $\vspace{0.05 in}\\
\hline\hline\\
\multicolumn{3}{c}{$Z=0.0002$}  \vspace{0.05 in}\\
\hline\\
\hspace{0.07 in}$ -3.205\pm0.020 $ & \hspace{0.09 in} $ 9.766\pm0.005 $ & \hspace{0.1 in} $\log M\leq 0.126 $\vspace{0.04 in}\\
\hspace{0.07 in}$ -2.646\pm0.018 $ & \hspace{0.09 in} $ 9.695\pm0.009 $ & \hspace{0.1 in} $ 0.126 <\log M\leq 0.368 $\vspace{0.04 in}\\
\hspace{0.07 in}$ -2.378\pm0.019 $ & \hspace{0.09 in} $ 9.596\pm0.014 $ & \hspace{0.1 in} $ 0.368 <\log M\leq 0.610 $\vspace{0.04 in}\\
\hspace{0.07 in}$ -2.005\pm0.021 $ & \hspace{0.09 in} $ 9.369\pm0.020 $ & \hspace{0.1 in} $ 0.610 <\log M\leq 0.852 $\vspace{0.04 in}\\
\hspace{0.07 in}$ -1.674\pm0.023 $ & \hspace{0.09 in} $ 9.087\pm0.028 $ & \hspace{0.1 in} $ 0.852 <\log M\leq 1.094 $\vspace{0.04 in}\\
\hspace{0.07 in}$ -1.249\pm0.026 $ & \hspace{0.09 in} $ 8.622\pm0.039 $ & \hspace{0.1 in} $ 1.094 <\log M\leq 1.336 $\vspace{0.04 in}\\
\hspace{0.07 in}$ -0.871\pm0.031 $ & \hspace{0.09 in} $ 8.116\pm0.053 $ & \hspace{0.1 in} $ 1.336 <\log M\leq 1.578 $\vspace{0.04 in}\\
\hspace{0.07 in}$ -0.598\pm0.037 $ & \hspace{0.09 in} $ 7.686\pm0.073 $ & \hspace{0.1 in} $\log M> 1.578 $\vspace{0.05 in}\\
\hline\hline\\
\multicolumn{3}{c}{$Z=0.0003$}  \vspace{0.05 in}\\
\hline\\
\hspace{0.07 in}$ -3.200\pm0.021 $ & \hspace{0.09 in} $ 9.772\pm0.005 $ & \hspace{0.1 in} $\log M\leq 0.128 $\vspace{0.04 in}\\
\hspace{0.07 in}$ -2.630\pm0.019 $ & \hspace{0.09 in} $ 9.699\pm0.009 $ & \hspace{0.1 in} $ 0.128 <\log M\leq 0.370 $\vspace{0.04 in}\\
\hspace{0.07 in}$ -2.393\pm0.020 $ & \hspace{0.09 in} $ 9.611\pm0.015 $ & \hspace{0.1 in} $ 0.370 <\log M\leq 0.612 $\vspace{0.04 in}\\
\hspace{0.07 in}$ -2.008\pm0.022 $ & \hspace{0.09 in} $ 9.376\pm0.021 $ & \hspace{0.1 in} $ 0.612 <\log M\leq 0.854 $\vspace{0.04 in}\\
\hspace{0.07 in}$ -1.682\pm0.024 $ & \hspace{0.09 in} $ 9.097\pm0.029 $ & \hspace{0.1 in} $ 0.854 <\log M\leq 1.095 $\vspace{0.04 in}\\
\hspace{0.07 in}$ -1.249\pm0.028 $ & \hspace{0.09 in} $ 8.624\pm0.040 $ & \hspace{0.1 in} $ 1.095 <\log M\leq 1.337 $\vspace{0.04 in}\\
\hspace{0.07 in}$ -0.870\pm0.032 $ & \hspace{0.09 in} $ 8.117\pm0.055 $ & \hspace{0.1 in} $ 1.337 <\log M\leq 1.579 $\vspace{0.04 in}\\
\hspace{0.07 in}$ -0.600\pm0.039 $ & \hspace{0.09 in} $ 7.690\pm0.076 $ & \hspace{0.1 in} $\log M> 1.579 $\vspace{0.05 in}\\
\hline\hline\\
\multicolumn{3}{c}{$Z=0.0004$}  \vspace{0.05 in}\\
\hline\\
\hspace{0.07 in}$ -3.201\pm0.022 $ & \hspace{0.09 in} $ 9.776\pm0.005 $ & \hspace{0.1 in} $\log M\leq 0.129 $\vspace{0.04 in}\\
\hspace{0.07 in}$ -2.616\pm0.020 $ & \hspace{0.09 in} $ 9.700\pm0.010 $ & \hspace{0.1 in} $ 0.129 <\log M\leq 0.371 $\vspace{0.04 in}\\
\hspace{0.07 in}$ -2.404\pm0.021 $ & \hspace{0.09 in} $ 9.622\pm0.016 $ & \hspace{0.1 in} $ 0.371 <\log M\leq 0.613 $\vspace{0.04 in}\\
\hspace{0.07 in}$ -2.006\pm0.023 $ & \hspace{0.09 in} $ 9.377\pm0.023 $ & \hspace{0.1 in} $ 0.613 <\log M\leq 0.855 $\vspace{0.04 in}\\
\hspace{0.07 in}$ -1.685\pm0.026 $ & \hspace{0.09 in} $ 9.103\pm0.031 $ & \hspace{0.1 in} $ 0.855 <\log M\leq 1.097 $\vspace{0.04 in}\\
\hspace{0.07 in}$ -1.249\pm0.029 $ & \hspace{0.09 in} $ 8.625\pm0.043 $ & \hspace{0.1 in} $ 1.097 <\log M\leq 1.339 $\vspace{0.04 in}\\
\hspace{0.07 in}$ -0.869\pm0.035 $ & \hspace{0.09 in} $ 8.116\pm0.059 $ & \hspace{0.1 in} $ 1.339 <\log M\leq 1.581 $\vspace{0.04 in}\\
\hspace{0.07 in}$ -0.595\pm0.042 $ & \hspace{0.09 in} $ 7.683\pm0.082 $ & \hspace{0.1 in} $\log M> 1.581 $\vspace{0.05 in}\\
\hline\hline\\
\end{tabular}
\begin{tabular}{ccr}
\\
\hline \hline\\
\hspace{0.05 in}$a$              & \hspace{0.09 in}$b$             & \hspace{0.1 in}validity range\vspace{0.05 in}\\
\hline\\
\multicolumn{3}{c}{$Z=0.0005$}  \vspace{0.05 in}\\
\hline\\
\hspace{0.07 in}$ -3.156\pm0.030 $ & \hspace{0.09 in} $ 9.773\pm0.006 $ & \hspace{0.1 in} $\log M\leq 0.112 $\vspace{0.04 in}\\
\hspace{0.07 in}$ -2.653\pm0.025 $ & \hspace{0.09 in} $ 9.717\pm0.012 $ & \hspace{0.1 in} $ 0.112 <\log M\leq 0.356 $\vspace{0.04 in}\\
\hspace{0.07 in}$ -2.428\pm0.026 $ & \hspace{0.09 in} $ 9.636\pm0.019 $ & \hspace{0.1 in} $ 0.356 <\log M\leq 0.601 $\vspace{0.04 in}\\
\hspace{0.07 in}$ -2.035\pm0.029 $ & \hspace{0.09 in} $ 9.400\pm0.028 $ & \hspace{0.1 in} $ 0.601 <\log M\leq 0.846 $\vspace{0.04 in}\\
\hspace{0.07 in}$ -1.697\pm0.032 $ & \hspace{0.09 in} $ 9.114\pm0.039 $ & \hspace{0.1 in} $ 0.846 <\log M\leq 1.090 $\vspace{0.04 in}\\
\hspace{0.07 in}$ -1.259\pm0.037 $ & \hspace{0.09 in} $ 8.637\pm0.054 $ & \hspace{0.1 in} $ 1.090 <\log M\leq 1.335 $\vspace{0.04 in}\\
\hspace{0.07 in}$ -0.869\pm0.043 $ & \hspace{0.09 in} $ 8.116\pm0.074 $ & \hspace{0.1 in} $ 1.335 <\log M\leq 1.579 $\vspace{0.04 in}\\
\hspace{0.07 in}$ -0.598\pm0.052 $ & \hspace{0.09 in} $ 7.688\pm0.102 $ & \hspace{0.1 in} $\log M> 1.579 $\vspace{0.05 in}\\
\hline\hline\\
\multicolumn{3}{c}{$Z=0.0006$} \vspace{0.05 in}\\
\hline\\
\hspace{0.07 in}$ -3.196\pm0.024 $ & \hspace{0.09 in} $ 9.785\pm0.006 $ & \hspace{0.1 in} $\log M\leq 0.132 $\vspace{0.04 in}\\
\hspace{0.07 in}$ -2.598\pm0.022 $ & \hspace{0.09 in} $ 9.706\pm0.011 $ & \hspace{0.1 in} $ 0.132 <\log M\leq 0.374 $\vspace{0.04 in}\\
\hspace{0.07 in}$ -2.428\pm0.023 $ & \hspace{0.09 in} $ 9.642\pm0.017 $ & \hspace{0.1 in} $ 0.374 <\log M\leq 0.616 $\vspace{0.04 in}\\
\hspace{0.07 in}$ -2.008\pm0.025 $ & \hspace{0.09 in} $ 9.383\pm0.024 $ & \hspace{0.1 in} $ 0.616 <\log M\leq 0.858 $\vspace{0.04 in}\\
\hspace{0.07 in}$ -1.677\pm0.027 $ & \hspace{0.09 in} $ 9.099\pm0.034 $ & \hspace{0.1 in} $ 0.858 <\log M\leq 1.100 $\vspace{0.04 in}\\
\hspace{0.07 in}$ -1.247\pm0.031 $ & \hspace{0.09 in} $ 8.626\pm0.046 $ & \hspace{0.1 in} $ 1.100 <\log M\leq 1.342 $\vspace{0.04 in}\\
\hspace{0.07 in}$ -0.867\pm0.037 $ & \hspace{0.09 in} $ 8.116\pm0.063 $ & \hspace{0.1 in} $ 1.342 <\log M\leq 1.584 $\vspace{0.04 in}\\
\hspace{0.07 in}$ -0.599\pm0.045 $ & \hspace{0.09 in} $ 7.692\pm0.087 $ & \hspace{0.1 in} $\log M> 1.584 $\vspace{0.05 in}\\
\hline\hline\\
\multicolumn{3}{c}{$Z=0.0008$} \vspace{0.05 in}\\
\hline\\
\hspace{0.07 in}$ -3.189\pm0.025 $ & \hspace{0.09 in} $ 9.792\pm0.006 $ & \hspace{0.1 in} $\log M\leq 0.134 $\vspace{0.04 in}\\
\hspace{0.07 in}$ -2.587\pm0.023 $ & \hspace{0.09 in} $ 9.712\pm0.011 $ & \hspace{0.1 in} $ 0.134 <\log M\leq 0.375 $\vspace{0.04 in}\\
\hspace{0.07 in}$ -2.453\pm0.024 $ & \hspace{0.09 in} $ 9.661\pm0.018 $ & \hspace{0.1 in} $ 0.375 <\log M\leq 0.617 $\vspace{0.04 in}\\
\hspace{0.07 in}$ -2.000\pm0.026 $ & \hspace{0.09 in} $ 9.382\pm0.026 $ & \hspace{0.1 in} $ 0.617 <\log M\leq 0.859 $\vspace{0.04 in}\\
\hspace{0.07 in}$ -1.687\pm0.029 $ & \hspace{0.09 in} $ 9.113\pm0.035 $ & \hspace{0.1 in} $ 0.859 <\log M\leq 1.101 $\vspace{0.04 in}\\
\hspace{0.07 in}$ -1.252\pm0.033 $ & \hspace{0.09 in} $ 8.634\pm0.048 $ & \hspace{0.1 in} $ 1.101 <\log M\leq 1.342 $\vspace{0.04 in}\\
\hspace{0.07 in}$ -0.866\pm0.038 $ & \hspace{0.09 in} $ 8.116\pm0.066 $ & \hspace{0.1 in} $ 1.342 <\log M\leq 1.584 $\vspace{0.04 in}\\
\hspace{0.07 in}$ -0.605\pm0.047 $ & \hspace{0.09 in} $ 7.702\pm0.091 $ & \hspace{0.1 in} $\log M> 1.584 $\vspace{0.05 in}\\
\hline\hline\\
\multicolumn{3}{c}{$Z=0.001$} \vspace{0.05 in}\\
\hline\\
\hspace{0.07 in}$ -3.178\pm0.030 $ & \hspace{0.09 in} $ 9.795\pm0.007 $ & \hspace{0.1 in} $\log M\leq 0.136 $\vspace{0.04 in}\\
\hspace{0.07 in}$ -2.567\pm0.028 $ & \hspace{0.09 in} $ 9.712\pm0.014 $ & \hspace{0.1 in} $ 0.136 <\log M\leq 0.377 $\vspace{0.04 in}\\
\hspace{0.07 in}$ -2.467\pm0.029 $ & \hspace{0.09 in} $ 9.674\pm0.022 $ & \hspace{0.1 in} $ 0.377 <\log M\leq 0.619 $\vspace{0.04 in}\\
\hspace{0.07 in}$ -2.018\pm0.032 $ & \hspace{0.09 in} $ 9.396\pm0.031 $ & \hspace{0.1 in} $ 0.619 <\log M\leq 0.860 $\vspace{0.04 in}\\
\hspace{0.07 in}$ -1.692\pm0.035 $ & \hspace{0.09 in} $ 9.116\pm0.043 $ & \hspace{0.1 in} $ 0.860 <\log M\leq 1.101 $\vspace{0.04 in}\\
\hspace{0.07 in}$ -1.248\pm0.040 $ & \hspace{0.09 in} $ 8.627\pm0.059 $ & \hspace{0.1 in} $ 1.101 <\log M\leq 1.343 $\vspace{0.04 in}\\
\hspace{0.07 in}$ -0.869\pm0.048 $ & \hspace{0.09 in} $ 8.119\pm0.081 $ & \hspace{0.1 in} $ 1.343 <\log M\leq 1.584 $\vspace{0.04 in}\\
\hspace{0.07 in}$ -0.599\pm0.058 $ & \hspace{0.09 in} $ 7.691\pm0.113 $ & \hspace{0.1 in} $\log M> 1.584 $\vspace{0.05 in}\\
\hline\hline\\
\end{tabular}
\end{table}

\begin{table}
\begin{tabular}{ccr}
\\
\hline \hline\\
\hspace{0.05 in}$a$              & \hspace{0.09 in}$b$             & \hspace{0.1 in}validity range\vspace{0.05 in}\\
\hline\\
\multicolumn{3}{c}{$Z=0.0012$}  \vspace{0.05 in}\\
\hline\\
\hspace{0.07 in}$ -3.189\pm0.028 $ & \hspace{0.09 in} $ 9.807\pm0.007 $ & \hspace{0.1 in} $\log M\leq 0.139 $\vspace{0.04 in}\\
\hspace{0.07 in}$ -2.566\pm0.025 $ & \hspace{0.09 in} $ 9.721\pm0.013 $ & \hspace{0.1 in} $ 0.139 <\log M\leq 0.381 $\vspace{0.04 in}\\
\hspace{0.07 in}$ -2.486\pm0.026 $ & \hspace{0.09 in} $ 9.690\pm0.020 $ & \hspace{0.1 in} $ 0.381 <\log M\leq 0.622 $\vspace{0.04 in}\\
\hspace{0.07 in}$ -2.007\pm0.029 $ & \hspace{0.09 in} $ 9.392\pm0.029 $ & \hspace{0.1 in} $ 0.622 <\log M\leq 0.864 $\vspace{0.04 in}\\
\hspace{0.07 in}$ -1.660\pm0.032 $ & \hspace{0.09 in} $ 9.092\pm0.040 $ & \hspace{0.1 in} $ 0.864 <\log M\leq 1.105 $\vspace{0.04 in}\\
\hspace{0.07 in}$ -1.242\pm0.037 $ & \hspace{0.09 in} $ 8.631\pm0.054 $ & \hspace{0.1 in} $ 1.105 <\log M\leq 1.347 $\vspace{0.04 in}\\
\hspace{0.07 in}$ -0.871\pm0.043 $ & \hspace{0.09 in} $ 8.131\pm0.074 $ & \hspace{0.1 in} $ 1.347 <\log M\leq 1.588 $\vspace{0.04 in}\\
\hspace{0.07 in}$ -0.623\pm0.052 $ & \hspace{0.09 in} $ 7.737\pm0.102 $ & \hspace{0.1 in} $\log M> 1.588 $\vspace{0.05 in}\\
\hline\hline\\
\multicolumn{3}{c}{$Z=0.0024$}  \vspace{0.05 in}\\
\hline\\
\hspace{0.07 in}$ -3.191\pm0.034 $ & \hspace{0.09 in} $ 9.848\pm0.009 $ & \hspace{0.1 in} $\log M\leq 0.150 $\vspace{0.04 in}\\
\hspace{0.07 in}$ -2.543\pm0.031 $ & \hspace{0.09 in} $ 9.751\pm0.016 $ & \hspace{0.1 in} $ 0.150 <\log M\leq 0.390 $\vspace{0.04 in}\\
\hspace{0.07 in}$ -2.570\pm0.032 $ & \hspace{0.09 in} $ 9.761\pm0.024 $ & \hspace{0.1 in} $ 0.390 <\log M\leq 0.629 $\vspace{0.04 in}\\
\hspace{0.07 in}$ -2.058\pm0.035 $ & \hspace{0.09 in} $ 9.439\pm0.035 $ & \hspace{0.1 in} $ 0.629 <\log M\leq 0.868 $\vspace{0.04 in}\\
\hspace{0.07 in}$ -1.663\pm0.040 $ & \hspace{0.09 in} $ 9.096\pm0.049 $ & \hspace{0.1 in} $ 0.868 <\log M\leq 1.108 $\vspace{0.04 in}\\
\hspace{0.07 in}$ -1.233\pm0.046 $ & \hspace{0.09 in} $ 8.620\pm0.067 $ & \hspace{0.1 in} $ 1.108 <\log M\leq 1.347 $\vspace{0.04 in}\\
\hspace{0.07 in}$ -0.890\pm0.053 $ & \hspace{0.09 in} $ 8.158\pm0.090 $ & \hspace{0.1 in} $ 1.347 <\log M\leq 1.587 $\vspace{0.04 in}\\
\hspace{0.07 in}$ -0.624\pm0.064 $ & \hspace{0.09 in} $ 7.736\pm0.125 $ & \hspace{0.1 in} $\log M> 1.587 $\vspace{0.05 in}\\
\hline\hline\\
\end{tabular}
\begin{tabular}{ccr}
\\
\hline \hline\\
\hspace{0.05 in}$a$              & \hspace{0.09 in}$b$             & \hspace{0.1 in}validity range\vspace{0.05 in}\\
\hline\\
\multicolumn{3}{c}{$Z=0.004$}  \vspace{0.05 in}\\
\hline\\
\hspace{0.07 in}$ -3.315\pm0.040 $ & \hspace{0.09 in} $ 9.840\pm0.010 $ & \hspace{0.1 in} $\log M\leq 0.148 $\vspace{0.04 in}\\
\hspace{0.07 in}$ -2.475\pm0.038 $ & \hspace{0.09 in} $ 9.715\pm0.018 $ & \hspace{0.1 in} $ 0.148 <\log M\leq 0.374 $\vspace{0.04 in}\\
\hspace{0.07 in}$ -2.703\pm0.039 $ & \hspace{0.09 in} $ 9.801\pm0.028 $ & \hspace{0.1 in} $ 0.374 <\log M\leq 0.600 $\vspace{0.04 in}\\
\hspace{0.07 in}$ -2.167\pm0.042 $ & \hspace{0.09 in} $ 9.479\pm0.040 $ & \hspace{0.1 in} $ 0.600 <\log M\leq 0.826 $\vspace{0.04 in}\\
\hspace{0.07 in}$ -1.812\pm0.047 $ & \hspace{0.09 in} $ 9.186\pm0.055 $ & \hspace{0.1 in} $ 0.826 <\log M\leq 1.053 $\vspace{0.04 in}\\
\hspace{0.07 in}$ -1.395\pm0.053 $ & \hspace{0.09 in} $ 8.746\pm0.075 $ & \hspace{0.1 in} $ 1.053 <\log M\leq 1.279 $\vspace{0.04 in}\\
\hspace{0.07 in}$ -0.956\pm0.064 $ & \hspace{0.09 in} $ 8.185\pm0.104 $ & \hspace{0.1 in} $ 1.279 <\log M\leq 1.505 $\vspace{0.04 in}\\
\hspace{0.07 in}$ -0.656\pm0.077 $ & \hspace{0.09 in} $ 7.734\pm0.143 $ & \hspace{0.1 in} $\log M> 1.505 $\vspace{0.05 in}\\
\hline\hline\\
\multicolumn{3}{c}{$Z=0.0085$}  \vspace{0.05 in}\\
\hline\\
\hspace{0.07 in}$ -3.287\pm0.040 $ & \hspace{0.09 in} $ 9.972\pm0.011 $ & \hspace{0.1 in} $\log M\leq 0.175 $\vspace{0.04 in}\\
\hspace{0.07 in}$ -2.467\pm0.036 $ & \hspace{0.09 in} $ 9.829\pm0.019 $ & \hspace{0.1 in} $ 0.175 <\log M\leq 0.399 $\vspace{0.04 in}\\
\hspace{0.07 in}$ -2.831\pm0.037 $ & \hspace{0.09 in} $ 9.974\pm0.027 $ & \hspace{0.1 in} $ 0.399 <\log M\leq 0.624 $\vspace{0.04 in}\\
\hspace{0.07 in}$ -2.232\pm0.040 $ & \hspace{0.09 in} $ 9.601\pm0.039 $ & \hspace{0.1 in} $ 0.624 <\log M\leq 0.849 $\vspace{0.04 in}\\
\hspace{0.07 in}$ -1.841\pm0.044 $ & \hspace{0.09 in} $ 9.269\pm0.053 $ & \hspace{0.1 in} $ 0.849 <\log M\leq 1.073 $\vspace{0.04 in}\\
\hspace{0.07 in}$ -1.348\pm0.051 $ & \hspace{0.09 in} $ 8.740\pm0.073 $ & \hspace{0.1 in} $ 1.073 <\log M\leq 1.298 $\vspace{0.04 in}\\
\hspace{0.07 in}$ -0.984\pm0.064 $ & \hspace{0.09 in} $ 8.268\pm0.105 $ & \hspace{0.1 in} $ 1.298 <\log M\leq 1.522 $\vspace{0.04 in}\\
\hspace{0.07 in}$ -0.790\pm0.076 $ & \hspace{0.09 in} $ 7.972\pm0.142 $ & \hspace{0.1 in} $\log M> 1.522 $\vspace{0.05 in}\\
\hline\\
\end{tabular}
\label{tab:taba2}
\end{table}

\begin{table}
\caption[]{Fitting equations of the relation between relative pulsation duration ($\delta t/t$ where $t$ is the age and $\delta t$ is the pulsation duration) and birth mass, $\log(\delta t/t)= D + \Sigma_{i=1}^4a _i\exp\left[-(\log M [{\rm M} _\odot]-b_i)^2/c_i^2\right]$.}
\begin{tabular}{ccccc}
\\
\hline\hline\\
\hspace{0.15 in} $D$ & \hspace{0.15 in}  $i$ & \hspace{0.20 in}$a$   & \hspace{0.20 in}$b$  & \hspace{0.20 in}$c$ \hspace{0.20 in} \vspace{0.05 in}  \\
\hline\\
\multicolumn{5}{c}{$Z=0.0001$} \vspace{0.05 in}  \\
\hline\\
\hspace{0.15 in}$-8.666$ &\hspace{0.15 in} 1 & \hspace{0.25 in} $ 2.678 $ &\hspace{0.25 in} $ 1.616 $ & \hspace{0.25 in} $ 0.308 $\hspace{0.20 in} \vspace{0.04 in}\\
                       &\hspace{0.15 in} 2 & \hspace{0.25 in} $ 5.534 $ &\hspace{0.25 in} $ 0.592 $ & \hspace{0.25 in} $ 1.020 $\hspace{0.20 in} \vspace{0.04 in}\\
                       &\hspace{0.15 in} 3 & \hspace{0.25 in} $ 0.220 $ &\hspace{0.25 in} $ 0.564 $ & \hspace{0.25 in} $ 0.084 $\hspace{0.20 in} \vspace{0.04 in}\\
                       &\hspace{0.15 in} 4 & \hspace{0.25 in} $ 4.532 $ &\hspace{0.25 in} $ 1.802 $ & \hspace{0.25 in} $ 0.109 $\hspace{0.20 in} \vspace{0.05 in}\\
\hline\hline\\
\multicolumn{5}{c}{$Z=0.0002$} \vspace{0.05 in}     \\
\hline\\
\hspace{0.15 in}$-4.693$ &\hspace{0.15 in} 1 & \hspace{0.25 in} $ 3.551 $ &\hspace{0.25 in} $ 1.696 $ & \hspace{0.25 in} $ 0.104 $\hspace{0.20 in} \vspace{0.04 in}\\
                       &\hspace{0.15 in} 2 & \hspace{0.25 in} $ 1.779 $ &\hspace{0.25 in} $ 0.587 $ & \hspace{0.25 in} $ 0.272 $\hspace{0.20 in} \vspace{0.04 in}\\
                       &\hspace{0.15 in} 3 & \hspace{0.25 in} $ 0.704 $ &\hspace{0.25 in} $ 0.272 $ & \hspace{0.25 in} $ 0.152 $\hspace{0.20 in} \vspace{0.04 in}\\
                       &\hspace{0.15 in} 4 & \hspace{0.25 in} $ 0.961 $ &\hspace{0.25 in} $ 1.371 $ & \hspace{0.25 in} $ 0.224 $\hspace{0.20 in} \vspace{0.05 in}\\
\hline\hline\\     
\multicolumn{5}{c}{$Z=0.0003$} \vspace{0.05 in}  \\
\hline\\
\hspace{0.15 in}$-5.977$ &\hspace{0.15 in} 1 & \hspace{0.25 in} $ 3.081 $ &\hspace{0.25 in} $ 0.608 $ & \hspace{0.25 in} $ 0.593 $\hspace{0.20 in} \vspace{0.04 in}\\
                       &\hspace{0.15 in} 2 & \hspace{0.25 in} $ 1.013 $ &\hspace{0.25 in} $ 1.283 $ & \hspace{0.25 in} $ 0.166 $\hspace{0.20 in} \vspace{0.04 in}\\
                       &\hspace{0.15 in} 3 & \hspace{0.25 in} $ 0.294 $ &\hspace{0.25 in} $ 0.263 $ & \hspace{0.25 in} $ 0.112 $\hspace{0.20 in} \vspace{0.04 in}\\
                       &\hspace{0.15 in} 4 & \hspace{0.25 in} $ 4.479 $ &\hspace{0.25 in} $ 1.788 $ & \hspace{0.25 in} $ 0.332 $\hspace{0.20 in} \vspace{0.05 in}\\
\hline\hline\\
\multicolumn{5}{c}{$Z=0.0004$} \vspace{0.05 in}  \\
\hline\\
\hspace{0.15 in}$-5.314$ &\hspace{0.15 in} 1 & \hspace{0.25 in} $ 5.276 $ &\hspace{0.25 in} $ 1.420 $ & \hspace{0.25 in} $ 1.069 $\hspace{0.20 in} \vspace{0.04 in}\\
                       &\hspace{0.15 in} 2 & \hspace{0.25 in} $ 0.293 $ &\hspace{0.25 in} $ 0.296 $ & \hspace{0.25 in} $ 0.079 $\hspace{0.20 in} \vspace{0.04 in}\\
                       &\hspace{0.15 in} 3 & \hspace{0.25 in} $ 13.226 $&\hspace{0.25 in}$ 1.222 $ & \hspace{0.25 in} $ 0.293 $\hspace{0.20 in} \vspace{0.04 in}\\
                       &\hspace{0.15 in} 4 & \hspace{0.25 in} $-16.215 $ &\hspace{0.25 in}$ 1.229 $ & \hspace{0.25 in} $ 0.348 $\hspace{0.20 in} \vspace{0.05 in}\\
\hline\hline\\
\multicolumn{5}{c}{$Z=0.0005$} \vspace{0.05 in}  \\
\hline\\
\hspace{0.15 in}$-7.132$ &\hspace{0.15 in} 1 & \hspace{0.25 in} $ 5.094 $ &\hspace{0.25 in} $ 1.689 $ & \hspace{0.25 in} $ 0.320 $\hspace{0.20 in} \vspace{0.04 in}\\
                       &\hspace{0.15 in} 2 & \hspace{0.25 in} $ -0.232 $ &\hspace{0.25 in} $ 0.430 $ & \hspace{0.25 in} $ 0.076 $\hspace{0.20 in} \vspace{0.04 in}\\
                       &\hspace{0.15 in} 3 & \hspace{0.25 in} $ 1.099 $ &\hspace{0.25 in} $ 1.220 $ & \hspace{0.25 in} $ 0.136 $\hspace{0.20 in} \vspace{0.04 in}\\
                       &\hspace{0.15 in} 4 & \hspace{0.25 in} $ 4.314 $ &\hspace{0.25 in} $ 0.553 $ & \hspace{0.25 in} $ 0.776 $\hspace{0.20 in} \vspace{0.05 in}\\
\hline\hline\\
\multicolumn{5}{c}{$Z=0.0006$} \vspace{0.05 in}  \\
\hline\\
\hspace{0.15 in}$-9.278$ &\hspace{0.15 in} 1 & \hspace{0.25 in} $ 6.180 $ &\hspace{0.25 in} $ 0.464 $ & \hspace{0.25 in} $ 0.913 $\hspace{0.20 in} \vspace{0.04 in}\\
                       &\hspace{0.15 in} 2 & \hspace{0.25 in} $ 0.502 $ &\hspace{0.25 in} $ 0.670 $ & \hspace{0.25 in} $ 0.150 $\hspace{0.20 in} \vspace{0.04 in}\\
                       &\hspace{0.15 in} 3 & \hspace{0.25 in} $ 1.964 $ &\hspace{0.25 in} $ 1.200 $ & \hspace{0.25 in} $ 0.195 $\hspace{0.20 in} \vspace{0.04 in}\\
                       &\hspace{0.15 in} 4 & \hspace{0.25 in} $ 6.911 $ &\hspace{0.25 in} $ 1.675 $ & \hspace{0.25 in} $ 0.306 $\hspace{0.20 in} \vspace{0.05 in}\\
\hline\hline\\
\end{tabular}
\begin{tabular}{ccccc}
\hline\hline\\
\hspace{0.15 in} D & \hspace{0.15 in}  $i$ & \hspace{0.20 in}$a$   & \hspace{0.20 in}$b$  & \hspace{0.20 in}$c$  \vspace{0.05 in}  \\
\hline\\   
\multicolumn{5}{c}{$Z=0.0008$} \vspace{0.05 in}  \\
\hline\\
\hspace{0.15 in}$-7.646$ &\hspace{0.15 in} 1 & \hspace{0.25 in} $ 1.223 $ &\hspace{0.25 in} $ 0.701 $ & \hspace{0.25 in} $ 0.193 $\hspace{0.20 in} \vspace{0.04 in}\\
                       &\hspace{0.15 in} 2 & \hspace{0.25 in} $ 4.447 $ &\hspace{0.25 in} $ 0.368 $ & \hspace{0.25 in} $ 0.621 $\hspace{0.20 in} \vspace{0.04 in}\\
                       &\hspace{0.15 in} 3 & \hspace{0.25 in} $ 1.403 $ &\hspace{0.25 in} $ 1.153 $ & \hspace{0.25 in} $ 0.115 $\hspace{0.20 in} \vspace{0.04 in}\\
                       &\hspace{0.15 in} 4 & \hspace{0.25 in} $ 6.283 $ &\hspace{0.25 in} $ 1.664 $ & \hspace{0.25 in} $ 0.461 $\hspace{0.20 in} \vspace{0.05 in}\\
\hline\hline\\
\multicolumn{5}{c}{$Z=0.001$} \vspace{0.05 in}  \\
\hline\\
\hspace{0.15 in}$-5.426$ &\hspace{0.15 in} 1 & \hspace{0.25 in} $ 1.607 $ &\hspace{0.25 in} $ 0.670 $ & \hspace{0.25 in} $ 0.206 $\hspace{0.20 in} \vspace{0.04 in}\\
                       &\hspace{0.15 in} 2 & \hspace{0.25 in} $ 1.542 $ &\hspace{0.25 in} $ 1.501 $ & \hspace{0.25 in} $ 0.112 $\hspace{0.20 in} \vspace{0.04 in}\\
                       &\hspace{0.15 in} 3 & \hspace{0.25 in} $ 2.944 $ &\hspace{0.25 in} $ 1.468 $ & \hspace{0.25 in} $ 0.475 $\hspace{0.20 in} \vspace{0.04 in}\\
                       &\hspace{0.15 in} 4 & \hspace{0.25 in} $ 2.177 $ &\hspace{0.25 in} $ 0.298 $ & \hspace{0.25 in} $ 0.364 $\hspace{0.20 in} \vspace{0.05 in}\\
\hline\hline\\
\multicolumn{5}{c}{$Z=0.0012$} \vspace{0.05 in}  \\
\hline\\
\hspace{0.15 in}$-6.014$ &\hspace{0.15 in} 1 & \hspace{0.25 in} $ 2.098 $ &\hspace{0.25 in} $ 0.728 $ & \hspace{0.25 in} $ 0.267 $\hspace{0.20 in} \vspace{0.04 in}\\
                       &\hspace{0.15 in} 2 & \hspace{0.25 in} $ 2.680 $ &\hspace{0.25 in} $ 0.279 $ & \hspace{0.25 in} $ 0.430 $\hspace{0.20 in} \vspace{0.04 in}\\
                       &\hspace{0.15 in} 3 & \hspace{0.25 in} $ 1.971 $ &\hspace{0.25 in} $ 1.181 $ & \hspace{0.25 in} $ 0.114 $\hspace{0.20 in} \vspace{0.04 in}\\
                       &\hspace{0.15 in} 4 & \hspace{0.25 in} $ 4.850 $ &\hspace{0.25 in} $ 1.532 $ & \hspace{0.25 in} $ 0.282 $\hspace{0.20 in} \vspace{0.05 in}\\
\hline\hline\\
\multicolumn{5}{c}{$Z=0.0024$} \vspace{0.05 in}  \\
\hline\\
\hspace{0.15 in}$-5.828$ &\hspace{0.15 in} 1 & \hspace{0.25 in} $ 2.799 $ &\hspace{0.25 in} $ 0.339 $ & \hspace{0.25 in} $ 0.478 $\hspace{0.20 in} \vspace{0.04 in}\\
                       &\hspace{0.15 in} 2 & \hspace{0.25 in} $ 3.797 $ &\hspace{0.25 in} $ 1.327 $ & \hspace{0.25 in} $ 0.334 $\hspace{0.20 in} \vspace{0.04 in}\\
                       &\hspace{0.15 in} 3 & \hspace{0.25 in} $ 2.412 $ &\hspace{0.25 in} $ 1.803 $ & \hspace{0.25 in} $ 0.150 $\hspace{0.20 in} \vspace{0.04 in}\\
                       &\hspace{0.15 in} 4 & \hspace{0.25 in} $ 1.263 $ &\hspace{0.25 in} $ 0.679 $ & \hspace{0.25 in} $ 0.167 $\hspace{0.20 in} \vspace{0.05 in}\\
\hline\hline\\
\multicolumn{5}{c}{$Z=0.004$} \vspace{0.05 in}  \\
\hline\\
\hspace{0.15 in}$-5.606$ &\hspace{0.15 in} 1 & \hspace{0.25 in} $ 0.991 $ &\hspace{0.25 in} $ 0.685 $ & \hspace{0.25 in} $ 0.156 $\hspace{0.20 in} \vspace{0.04 in}\\
                       &\hspace{0.15 in} 2 & \hspace{0.25 in} $ 2.719 $ &\hspace{0.25 in} $ 1.712 $ & \hspace{0.25 in} $ 0.163 $\hspace{0.20 in} \vspace{0.04 in}\\
                       &\hspace{0.15 in} 3 & \hspace{0.25 in} $ 2.670 $ &\hspace{0.25 in} $ 0.357 $ & \hspace{0.25 in} $ 0.494 $\hspace{0.20 in} \vspace{0.04 in}\\
                       &\hspace{0.15 in} 4 & \hspace{0.25 in} $ 3.806 $ &\hspace{0.25 in} $ 1.255 $ & \hspace{0.25 in} $ 0.309 $\hspace{0.20 in} \vspace{0.05 in}\\
\hline\hline\\
\multicolumn{5}{c}{$Z=0.0085$} \vspace{0.05 in}  \\
\hline\\
\hspace{0.15 in}$-4.754$ &\hspace{0.15 in} 1 & \hspace{0.25 in} $ 1.002 $ &\hspace{0.25 in} $ 0.682 $ & \hspace{0.25 in} $ 0.116 $\hspace{0.20 in} \vspace{0.04 in}\\
                       &\hspace{0.15 in} 2 & \hspace{0.25 in} $ 2.031 $ &\hspace{0.25 in} $ 0.381 $ & \hspace{0.25 in} $ 0.343 $\hspace{0.20 in} \vspace{0.04 in}\\
                       &\hspace{0.15 in} 3 & \hspace{0.25 in} $ 3.306 $ &\hspace{0.25 in} $ 1.327 $ & \hspace{0.25 in} $ 0.400 $\hspace{0.20 in} \vspace{0.04 in}\\
                       &\hspace{0.15 in} 4 & \hspace{0.25 in} $ 94.651$ &\hspace{0.25 in} $ 1.860 $ & \hspace{0.25 in} $ 0.060 $\hspace{0.20 in} \vspace{0.05 in}\\
\hline
\end{tabular}
\end{table}

\label{lastpage}
\end{document}